\documentclass[linenumbers]{aastex631}

\newcommand{\msolar}{M$_{\odot}$}
\newcommand{\lsolar}{L$_{\odot}$}
\usepackage{CJK}
\usepackage[whole]{bxcjkjatype} 
\usepackage{CJKutf8}

\begin{document}
\nolinenumbers

\title{The environmental dependence of {\it Spitzer} dusty Supernovae}

\author[0000-0002-6986-5593]{Lin Xiao}
\altaffiliation{corresponding author: linxiao@hbu.edu.cn}
\affiliation{Department of Physics, Hebei University, Baoding, 071002, China}
\affiliation{Hebei Key Laboratory of High-precision Computation and Application of Quantum Field Theory, Baoding, 071002, China}
\affiliation{Hebei Research Center of the Basic Discipline for Computational Physics, Baoding, 071002, China}

\author[0000-0003-4610-1117]{Tam\'as Szalai}
\affiliation{Department of Experimental Physics, Institute of Physics, University of Szeged, H-6720 Szeged, Dóm tér 9, Hungary}
\affiliation{HUN-REN--SZTE Stellar Astrophysics Research Group, H-6500 Baja, Szegedi {\'u}t, Kt. 766, Hungary}

\author[0000-0002-1296-6887]{Llu\'is Galbany}
\affiliation{Institute of Space Sciences (ICE, CSIC), Campus UAB, Carrer de Can Magrans, s/n, E-08193 Barcelona, Spain}
\affiliation{Institut d’Estudis Espacials de Catalunya (IEEC), E-08034 Barcelona, Spain}

%\collaboration{20}{(AAS Journals Data Editors)}
\author[0000-0003-2238-1572]{Ori Fox}
\affiliation{Space Telescope Science Institute, 3700 San Martin Drive, Baltimore, MD 21218, USA}

\author[0000-0001-7201-1938]{Lei Hu}
%\affiliation{Purple Mountain Observatory Nanjing 210023, People’s Republic of China}
\affiliation{McWilliams Center for Cosmology, Department of Physics, Carnegie Mellon University, 5000 Forbes Ave, Pittsburgh, 15213, PA, USA}

\author[0000-0003-3031-6105]{Maokai Hu}
\affiliation{Purple Mountain Observatory, Chinese Academy of Sciences, Nanjing 210023, China}

\author[0000-0002-6535-8500]{Yi Yang
\begin{CJK}{UTF8}{gbsn}
(杨轶)
\end{CJK}}
\affiliation{Physics Department and Tsinghua Center for Astrophysics (THCA), Tsinghua University, Beijing, 100084, China}
\affiliation{Department of Astronomy, University of California, Berkeley, CA 94720-3411, USA}

\author[0000-0003-1169-1954]{Takashi J. Moriya}
\affiliation{National Astronomical Observatory of Japan, National Institutes of Natural Sciences, 2-21-1 Osawa, Mitaka, Tokyo 181-8588, Japan}
\affiliation{Graduate Institute for Advanced Studies, SOKENDAI, 2-21-1 Osawa, Mitaka, Tokyo 181-8588, Japan}
\affiliation{School of Physics and Astronomy, Monash University, Clayton, Victoria 3800, Australia}

\author[0000-0001-6540-0767]{Thallis Pessi}
\affil{Instituto de Estudios Astrof\'isicos, Facultad de Ingenier\'ia y Ciencias, Universidad Diego Portales, Av. Ej\'ercito Libertador 441, Santiago, Chile}

\author[0000-0001-9204-7778]{Zhanwen Han}
\affiliation{Yunnan Observatories, Chinese Academy of Sciences, Kunming 650216, China}
\affiliation{Key Laboratory for the Structure and Evolution of Celestial Objects, Yunnan Observatories, CAS, Kunming 650216, China}
\affiliation{International Centre of Supernovae, Yunnan Key Laboratory, Kunming, 650216, China}

\author[0000-0002-7334-2357]{Xiaofeng Wang}
\affiliation{Physics Department and Tsinghua Center for Astrophysics (THCA), Tsinghua University, Beijing, 100084, China}
\affiliation{Beijing Planetarium, Beijing Academy of Science and Technology, Beijing, 100044, China}

\author[0009-0004-4256-1209]{Shengyu Yan}
\affiliation{Physics Department and Tsinghua Center for Astrophysics (THCA), Tsinghua University, Beijing, 100084, China}



\begin{abstract}
\nolinenumbers
Thanks to the mid-infrared capability offered by {\it Spitzer}, systematic searches of dust in SNe have been carried out over the past decade. Studies have revealed the presence of a substantial amount of dust over a broad range of SN subtypes. How normal SNe present mid-IR excess at later time and turn out to be dusty SNe can be affected by several factors, such as mass-loss history and envelope structure of progenitors and their explosion environment. All these can be combined and related to their environmental properties. A systematic analysis of SNe that exploded under a dusty environment could be of critical importance to measure the properties of the dust-veiled exploding stars, and whether such an intense dust production process is associated with the local environment. In this work, we firstly use the IFS data to study the environmental properties of dusty SNe compared to those of normal ones, and analyze correlations between the environmental properties and their dust parameters. We find that dusty SNe have a larger proportion located at higher SFR regions compared to the normal types. The occurrence of dusty SNe is less dependent on metallicity, with the oxygen abundance spanning from subsolar to oversolar metallicity. We also find the host extinction of dusty SNe scatters a lot, with about 40\% of dusty SN located at extremely low extinction environments, and another 30\% of them with considerably high host extinction of E(B-V)$>$0.6 mag.

\end{abstract}



\section{Introduction} \label{sec:intro}

The studies of the local environments where supernovae (SNe) explode provide compelling measurements of various host properties, such as metallicity, extinction, star formation rate (SFR), and the age of the SN parent population. These studies consequently have achieved a lot of important results, revealing an irreplaceable supplement to provide independent constraints on the properties of their progenitor systems \citep{Anderson2015,Chen2017,Galbany2018,Kuncarayakti2018,Xiao2019}; the environmental dependence of SN occurrence, optical light curve properties and mass-loss histories \citep{Taddia2015,Graur2017,Galbany2017,Moriya2023,Pessi2023}; the connection between star formation to death which puts SNe on the role of galaxy studies \citep{Crowther2013,Galbany2014,Galbany2016,Xiao2015,Gupta2016,Chruslinska2019}. 
 
The environmental study of SNe are more powerful to statistically analyze different types of SN populations. Various comparative studies of different SN types provide us another perspective on the diversity and correlation of different SN types. \cite{Galbany2018} and \cite{Pessi2023} separately apply statistical test to quantitatively analyze and provide a statistical measurement on associations of different SN types in terms of different environmental properties. Through the study of the parent stellar populations at the SN explosion sites, \cite{Kuncarayakti2018} and \cite{Xiao2019} also find the initial masses of hydrogen-rich and hydrogen-poor SN progenitor are statistically similar which is considered as the result of binary-star evolution. Additionally, the correlation measurement between the observed light-curve parameters and local environmental properties of SNe become possible. \cite{Galbany2018} found SN II post-maximum brightness decline with the local SFR intensity, corresponding to the range of decline rates of fast-declining (IIL) and slow-declining (IIP) SN subtypes. \cite{Moriya2023} studied the environmental dependence of type IIn SNe on the light curve features and their progenitor mass-loss rate. They found that SNe IIn with higher peak luminosity tend to occur in low-metallicity environments, but the circumstellar medium (CSM) density related with mass-loss rates of progenitors is not significantly metallicity dependent. 

Thanks to the mid infrared (mid-IR) capability offered by the {\it Spitzer Space Telescope} ($Spitzer$ hereafter), systematic searches of dust in SNe have been carried out over the past decade (see, \citealp{Szalai2019,Szalai2021} for reviews). Most of the dust emission in SNe can be attributed to being either pre-existing or newly formed during later-time CSM interactions \citep{Fox2011, Szalai2013}. The inferred dust mass is critical to determine whether SN is playing a major role in enriching the dust budget within the interstellar medium (ISM). The dust grain species and the radial distribution profile would provide an immediate trace of the progenitor's pre-explosion mass-loss history, thus constraining the progenitor system \citep{Matsuura2009,Fox2011}. Characterizations of the dust present in such dusty SNe have revealed the presence of a substantial amount of dust over a broad range of SN subtypes. \cite{Szalai2019} find that type IIP SNe make up nearly 20\% of all detected dusty SNe which is very close to the percentage of IIn with 21\%. Additionally, even the stripped-envelope SNe (IIb,Ib and Ic), can be an important dust producer taking account for another 20\%. Type Ia including its subtypes can also be mid-IR bright with significant dust component and contribute to 20\% of total detected dusty SNe in Spitzer. 

The question of what kind of SNe can be dusty and others are not, is a critical issue to resolve with the forthcoming mid-IR survey era after the launch of {\it James Webb Space Telescope} ({\it JWST}). The previous studies of dusty SNe by $Spitzer$ and {\it JWST} are more focused on the measurement of dust components, the formation of dust, and the contribution of dust emission, in either individual or statistical SN studies, and do not answer the question. How normal SNe can become dusty SNe at later time can be affected by several factors: the mass-loss history and the envelope structure of progenitors, and their explosion environment. All these factors can be combined and related to the properties of their explosion environment. A systematic characterization of SNe that exploded under a dusty environment, could be of critical importance to measure the properties of the dust-veiled exploding stars, and whether such an intense dust production process is associated with the local environment of the progenitor population.

Therefore, in this work, we firstly try to find a method to answer the question using the environment study of dusty SNe and compared to those normal counterparts. We expect that the difference can give us a hint to answer the question. The outline of this paper is described as follows. We introduce the selection of our dusty SN sample in Section \ref{sec:sample}. Section \ref{sec:result} presents the statistical result of the comparative study of dusty SNe with the normal ones in terms of four characteristic environmental properties, which are the H$\alpha$ equivalent width (EW), the star formation rate intensity log$\Sigma_{\rm SFR}$, the metallicity in terms of oxygen abundance 12+log(O/H)$_{\rm D16}$ measured by \cite{Dopita2016}, and the extinction of each SN host HII regions. In Section \ref{sec:dependence}, we investigate if there exists any correlation between the SN environmental properties and their dust parameters. Finally, we conclude this paper in Section \ref{sec:conclusions}.\\

% --Have you ever tried to plot some of these parameter pairs for all SNe (e.g., SFR vs Oxygen abundance), and see if the dusty ones are grouped, or distinct from other subclasses?  Lin: Thanks, YY. I may think about this.

% --I'm not sure how easy your routines can derive those parameters. Probably consider doing the local vs global plot for some parameters, and see if different subtypes, and the dusty SNe, will group together; 

%--After that, you may want to see if any correlations can be found between the properties of the local environment and newly formed dust. For each dust parameter, it may also be possible to define a measurement at a certain phase, if it would be feasible; \\

% --The sample size appears to be small enough to probe any redshift evolution. For large samples, this can also be done for various parameters.  Lin: This can be thought about in the WISE sample in prep. 

\section{Sample selection} \label{sec:sample}

We compile a sample of dusty SNe based on all SNe observed by {\it Spitzer} as reported in Tables 4 and 1 of \citet{Szalai2019} and \citet{Szalai2021}, respectively. Among a total number of 1142 observed SN sites, 151 events with at least one epoch of positive detection in the {\it Spitzer}/IRAC data were identified as `dusty SNe', spanning various subtypes from thermonuclear to core-collapse SNe. We remark that some of them were only observed at a single epoch at $\lesssim$50 days after the explosion. At such early phases, the temperature is still too high for condensation to take place and form dust grains. The mid-IR emission is still dominated by the hot gas in the SN ejecta, measurement of dust properties is thus not feasible. Despite the confusion introduced by recognizing these mid-IR bright SNe as `dusty', considering the limited number of such cases and the lack of understanding of the exact emission sources in the mid-IR before case-by-case modeling, we still include them in our sample of dusty SNe. 

%\textcolor{blue}{\bf YY: This paragraph is too long. Break it down.}
During the first year after SN explosion, emissions from the hot SN ejecta can still contribute a substantial amount of flux observed in the mid-IR, thus contaminating the thermal emissions by either or both the dust in pre-existing CSM or newly-formed dust grains due to later-time CSM interactions. As the ejecta cools and the forward shock expands, after the first year, the mid-IR emission becomes progressively dominated by the thermal emission from the newly-formed dust \citep{Fox2011,Szalai2013}. We, therefore, estimate the dust properties based on the observations and fittings between 400 and 600 days to minimize the contamination from the SN ejecta. The dust properties were derived following the prescription described by \citet{Szalai2019}. This time scale is also the most likely to have dust observed in literature.

For SN 2014J, we use the image from Wide-field Infrared Survey Explorer ({\it WISE}) to extend their mid-IR detection to allow dust produced and measured at later time. SN 2014J is an unusual case of Ia, its luminous mid-IR emission can be affected by extinction and dust scattering from both CSM and ISM according to analysis of the polarization signal occurring 277 day after maximum light in \cite{Yang2018}. The measurement of dust based on {\it WISE} at 280 day provide a dust mass $\sim$ $10^{-6}$\msolar, which is consistent with result from polarization. Because there are no good detection of SN 2014J in WISE after 400 days, we use the dust measurement at 280 day for it in this work. The mid-IR photometry of SN 2014J from WISE are presented in Section \ref{sec:2014J-WISE} 

For other SNe lack of both {\it Spitzer} and {\it WISE} mid-IR observations between 400 and 600 days, we infer the estimated dust temperature (Figure~\ref{fig:inter_Tdust}), mass (Figure~\ref{fig:inter_Mdust}) and luminosity (Figure~\ref{fig:inter_Ldust}) by a linear regression to the dust parameters summarised in \cite{Szalai2019}. More details for the process of the linear regression can be found in Section \ref{sec:linear_dust}. We consider the fitting results for any particular subtype of SNe plausible when a criterion of regression $p<0.5$ is reached. Then, we achieved the dust properties for the cases of SNe 2004G, 2005ip, 2006jd, 2007gr and 2012cd presented in Table~\ref{tab:sndata} where mid-IR data at later phases are missing. We note that parameters of newly-formed dust at different phases are heavily dependent on the radial density profiles and the mass of the ejecta (see, a comprehensive discussion by \citealp{Sarangi2022b}), but the adopted linear regression is sufficient for investigating the dust properties of the sample of dusty SNe. 

On the other side, to study the SN environment we collected 272 SNe including all SN types from \cite{Galbany2018} (hereafter G18), 111 CCSNe from \cite{Pessi2023} (hereafter P23) and 21 IIn SNe from \cite{Moriya2023} (hereafter M23). As a consequence, we achieved 395 SN sites observed by Integrated Field spectroscopy (IFS) excluding the duplicate SNe in the three samples. The compilation provides by far the largest and most diverse sample of SN explosion site with IFS observations in literature. Furthermore, we divided the 395 SNe into four groups, with 118 Ia SNe, 154 hydrogen-rich SNe (IIP/IIL/II), 83 hydrogen-poor SNe (IIb/Ib/Ic), and 38 IIn SNe. We note our sample was taken from different works using different facilities. Firstly, G18 sample is constructed using 232 observed galaxies with IFS data from the the PMAS/PPak Integral field Supernova hosts COmpilation (PISCO; \citealp{Galbany2018}), which is updated in terms of different SN subtypes including SN Ia and increased in the completeness of SN host galaxy sample with low-mass ($<10^{10}$\msolar) galaxies. So the host galaxies have $M_r$ varied from -24 to -12 mag and redshift up to 0.09 with median $\sim$ 0.017. Due to the spatial pixel (spaxel) size of PISCO at $1^{"}\times1^{"}$ (namely 1 arcsec$^{2}$ in area), a typical size of the HII region segregation in PISCO data can be a few hundreds of parsecs depending on distance \citep{Galbany2018}. This size is significantly larger than individual HII regions, and usually 1 to 6 HII regions were selected in data with that resolution \citep{Mast2014}. 

Then for P23 sample, they are selected from the SN catalogue of All-Sky Automated Survey for Supernovae (ASAS-SN) and then covered by the All-weather MUse Supernova Integral field Nearby Galaxies (AMUSING) survey \citep{Galbany2016a}. Their host galaxies have redshift up to 0.06 with median $\sim$ 0.011 and $M_B$ ranging from -23 to -16 mag. For AMUSING data, the spaxel size is $0.2^{"}\times0.2^{"}$, five times smaller than PISCO, which can zoom into SN parent stellar population better and decrease the contamination from nearby ones. M23 sample selected only type IIn host galaxies from the PISCO, AMUSING, and the Mapping Nearby Galaxies at APO (MaNGA; \citealp{Bundy2015}) surveys. The spaxel size of MaNGA is $0.5^{"}\times0.5^{"}$, in the middle of that of PISCO and AMUSING. As a whole, the HII region segmentation under the resolution of the three samples is larger than the physical scale of a real HII region, more or less, and the contamination from nearby stellar populations would affect our result using it as a SN parent stellar population. Compared to that the dust radius measured by blackbody fit is in the order of $10^{16}$cm (just about 1\% of parsec; \citealt{Szalai2019}), the IFS data with a much larger size cannot touch on the close environment of SNe-CSM interaction region. However, the studies based on IFS data do link the star-forming environment where those SN progenitors were born to their death end. Our IFS sample are a collection of these three samples, and G18 sample makes up nearly 70\% of the overall sample which show a more spread distribution in host galaxy magnitude and redshift. This makes sure the completeness of our sample in terms of galaxy brightness and size. P23 sample is smaller than G18 sample and enriches the whole sample with more CCSNe and new SNe detected after 2017. M23 sample is the smallest part with extra SNe IIn. In general, our IFS sample can provide an unbiased analysis of different types of SN environments. 

Finally, we cross-matched the samples of 1) the {\it Spitzer} dusty SNe and 2) the PISCO/AMUSING/MaNGA SNe local environment and found 9 SNe with data obtained in both mid-IR and IFS. In Table~\ref{tab:sndata} we present their dust properties, including the temperature, mass, and luminosity of the dust as derived based on direct observations or fitting models. Table \ref{tab:Environmentdata} reports the local environmental properties of the cross-matched sample of 9 SNe. We measured the EW(H$\alpha$) which can be a good measurement of the stellar population age of SN host regions, the star formation rate intensity log$\Sigma_{\rm SFR}$ to trace the relation between star forming and SN explosion, the metallicity indicated by the oxygen abundance 12+log(O/H)$_{\rm D16}$ measured by \cite{Dopita2016}, and the E(B-V) of the SN host HII regions to recognize any significant effect of extinction caused by ambient CSM or ISM, respectively. A more detailed description of the calculation of the SN local environmental properties can be found in \citet{Galbany2018}.

\begin{deluxetable}{cccccccc}
\tablecaption{The SN sample with both {\it Spitzer/WISE} mid-IR and IFS detection used in this work and its dust properties. \label{tab:sndata}}
% \tablenum{1}
\tablecolumns{3}
\tablehead{
\colhead{Name} & \colhead{Type} & \colhead{Distance} & \colhead{Epoch} & \colhead{T$_{\rm dust}$} & \colhead{M$_{\rm dust}$} & \colhead{L$_{\rm dust} $} & \colhead{References}\\
\colhead{ } & \colhead{ } & \colhead{Mpc} & \colhead{days} & \colhead{K} & \colhead{$10^{-5}$\msolar} & \colhead{$10^{6}$\lsolar }
}
\startdata
 \tableline
SN2003gd & IIP & 8.9$\pm$3.2 & 409 & 520 & 1.1 & 0.1 & 1 \\
SN2004G & II & 26.9$\pm$6.9 & \textcolor{gray}{500} & - & \textcolor{gray}{57} & \textcolor{gray}{1.0} & 2\\ 
SN2005ip & IIn & 30.0$\pm$7.2 & \textcolor{gray}{450} & \textcolor{gray}{682} & \textcolor{gray}{805} & \textcolor{gray}{452} &  2,3,4\\
SN2006jd & IIn & 77.0$\pm$5.0 & \textcolor{gray}{500} & \textcolor{gray}{675} & \textcolor{gray}{833} & \textcolor{gray}{466} &  2,4,5\\
SN2007gr & Ic & 9.3$\pm$1.2 & \textcolor{gray}{550} & \textcolor{gray}{553} & \textcolor{gray}{5.2} & \textcolor{gray}{0.3} &  6 \\
SN2012cd & IIb & 50.0$\pm$3.7 & \textcolor{gray}{450} & \textcolor{gray}{548} & \textcolor{gray}{8.5} & \textcolor{gray}{0.5} &  2 \\
SN2013ej & IIP/L & 9.5$\pm$0.6 & 439 & 360 &  64 & 0.4 & 2,7 \\
SN2014J & Ia & 3.3$\pm$0.2 &  280 &  908 & 0.4 &  0.1 &  8,9\\
SN2017hcc & IIn & 72.0$\pm$6.0 &  565 &  950 &  140 &  270 &  10\\
\tableline
\enddata
\tablecomments{The gray numbers mean the dust properties are derived from linear regression as discussed in Section \ref{sec:linear_dust}, not from direct observations. \\
References. (1)\cite{Meikle2007}, (2)\cite{Szalai2019}, (3)\cite{Fox2010}, (4)\cite{Fox2013}, (5)\cite{Stritzinger2012}, (6)\cite{Kochanek2011}, (7)\cite{Mauerhan2017},(8)\cite{Johansson2017},(9)\cite{Yang2018},(10)\cite{Szalai2021}}
\end{deluxetable}

\begin{deluxetable}{cccccc}
\tablecaption{The local environmental properties of the 9 SNe with both {\it Spitzer} mid-IR and IFS detection used in this work. 
\label{tab:Environmentdata}}
% \tablenum{1}

\tablecolumns{5}
\tablehead{
\colhead{Name} & \colhead{Host} & \colhead{EW(H$\alpha$)} & \colhead{12+log(O/H)$_{\rm D16}$} & \colhead{log$\Sigma_{\rm SFR}$} & \colhead{E(B-V)} \\
\colhead{ } & \colhead{ } & \colhead{\AA} & \colhead{ } & \colhead{\msolar yr$^{-1}$kpc$^{-2}$} & \colhead{mag} 
}
\startdata
 \tableline
SN2003gd	&	NGC 628	&	43.18	$\pm$	3.01	&	8.49	$\pm$	0.01	&	-2.241 	$\pm$	0.021 	&	0.003 	$\pm$	0.121 	\\
SN2004G	&	NGC 5668	&	-			&	8.22	$\pm$	0.01	&	0.802 	$\pm$	0.021 	&	0.000 			\\
SN2005ip	&	NGC 2906	&	34.87	$\pm$	1.45	&	8.82	$\pm$	0.03	&	-1.310 	$\pm$	0.025 	&	0.412 	$\pm$	0.098 	\\
SN2006jd	&	UGC 4179	&	-			&	8.26	$\pm$	0.01	&	-2.260 	$\pm$	0.027 	&	0.721 	$\pm$	0.213 	\\
SN2007gr	&	NGC 1058	&	16.19	$\pm$	0.49	&	8.63	$\pm$	0.01	&	-2.120 	$\pm$	0.022 	&	0.060 	$\pm$	0.043 	\\
SN2012cd	&	MCG +09-22-53	&	-			&	8.39	$\pm$	0.01	&	-2.556 	$\pm$	0.034 	&	0.000 	$\pm$	0.102 	\\
SN2013ej	&	NGC 628	&	17.59	$\pm$	1.20	&	8.73	$\pm$	0.02	&	1.065 	$\pm$	0.021 	&	1.008 	$\pm$	0.237 	\\
SN2014J	&	NGC 3034	&	22.18	$\pm$	0.33	&	8.91	$\pm$	0.03	&	-0.445 	$\pm$	0.030 	&	0.886 	$\pm$	0.052 	\\
SN2017hcc	&	anon.	&	13.44	$\pm$	0.59	&	8.06	$\pm$	0.33	&	-4.192 	$\pm$	0.000 	&	0.200 	$\pm$	0.030 	\\
\tableline
\enddata
\end{deluxetable}

\section{Statistical Result} \label{sec:result}
%The aim of this work is to collate the {\it Spitzer} dusty SNe and then compare to literature SN host environmental properties with the median and mean value listed in Table \ref{tab:median4types}. 
In this section, we outline the statistical result of the comparative study in terms of four characteristic environmental properties, which are the EW(H$\alpha$), the star formation rate intensity log$\Sigma_{\rm SFR}$, the metallicity in terms of oxygen abundance 12+log(O/H)$_{\rm D16}$ measured by \cite{Dopita2016}, and the SN host extinction E(B-V). We note our dusty SN sample include both Ia and CCSNe. Given only one Ia in the sample, we decided to pick out the Ia SN from the sample and created a group of dusty CCSNe for the comparison with normal types in distributions of cumulative fraction of SNe and Kolmogorov–Smirnov (KS) statistic matrix. The median and mean value of the environmental properties of dusty SNe including Ia and dusty CCSNe are listed in Table \ref{tab:median4types}.

\begin{deluxetable}{ccccc}
\tablecaption{The median and mean value of the environmental properties for four SN subtypes and the dusty SN sample in Table \ref{tab:Environmentdata}. The label of 'Dusty SNe' is for all the 9 selected SNe including Ia, and the label of 'Dusty CCSNe' is for the 8 CCSNe with the Ia SN removed.
\label{tab:median4types}}
% \tablenum{1}
\tablecolumns{3}
\tablehead{
\colhead{Type} & \colhead{logEW(H$\alpha$)} & \colhead{12+log(O/H)$_{\rm D16}$} & \colhead{log$\Sigma_{\rm SFR}$} & \colhead{E(B-V)} \\
\colhead{ } & \colhead{\rm \AA} & \colhead{ } & \colhead{\msolar yr$^{-1}$kpc$^{-2}$} & \colhead{mag} 
}
\startdata
 \tableline
& & Median & & \\
\cline{2-5}
H-rich & 1.65 & 8.45 & -1.93 & 0.19 \\
H-poor & 1.64 & 8.54 & -1.87 & 0.15 \\ 
IIn & 1.60 &  8.60 & -2.34 & 0.20 \\
Ia & 1.10 & 8.68 & -2.36 & 0.10 \\
\textbf{Dusty SNe} & \textbf{1.30} & \textbf{8.49} & \textbf{-2.12} & \textbf{0.20} \\
\textbf{Dusty CCSNe} & \textbf{1.25} & \textbf{8.44} & \textbf{-2.18} & \textbf{0.20} \\
\tableline
& & Mean & & \\
\cline{2-5}
H-rich & 1.67$\pm$0.50 & 8.43$\pm$0.34 & -1.86$\pm$0.93 & 0.21$\pm$0.19 \\
H-poor & 1.58$\pm$0.45 & 8.49$\pm$0.33 & -1.83$\pm$0.73 & 0.19$\pm$0.17 \\ 
IIn & 1.60$\pm$0.40 &  8.52$\pm$0.28 & -2.61$\pm$1.20 & 0.23$\pm$0.16 \\
Ia & 0.81$\pm$0.80 & 8.66$\pm$0.21 & -2.34$\pm$0.79 & 0.17$\pm$0.23 \\
\textbf{Dusty SNe} & \textbf{1.35$\pm$0.19} & \textbf{8.50$\pm$0.29} & \textbf{-1.47$\pm$1.60} & \textbf{0.36$\pm$0.41} \\
\textbf{Dusty CCSNe} & \textbf{1.35$\pm$0.21} & \textbf{8.45$\pm$0.27} & \textbf{-1.60$\pm$1.76} & \textbf{0.30$\pm$0.38} \\
\tableline
\enddata
\end{deluxetable}

\subsection{H$\alpha$ equivalent width} \label{subsec:EW_Ha}

\begin{figure}[ht!]
\centering
\includegraphics[width=0.49\columnwidth]{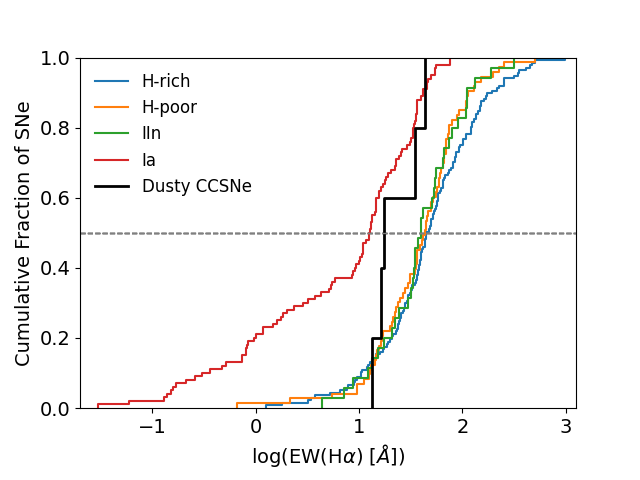}
\includegraphics[width=0.49\columnwidth]{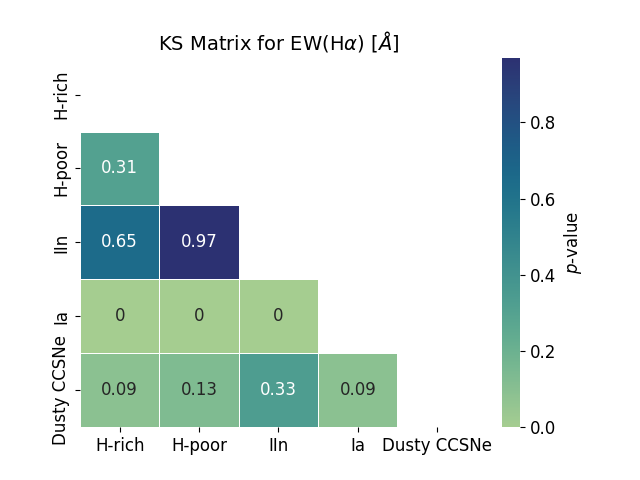}\\
\caption{Left: normalized cumulative distributions of H$\alpha$ equivalent width of SNe hosted HII regions in four types: H-rich in blue line, H-poor in orange line, IIn in green line and Ia in red line. A dotted horizontal line at 0.5 fraction represents the median value of the distributions. Right: KS matrix for each combination of different SN types. 
\label{fig:hist_EW}}
\end{figure}

Figure \ref{fig:hist_EW} shows the cumulative distributions of EW(H$\alpha$) of the SN-hosted HII regions for different SN types, and the KS test for each combination of them. A similar distribution is shared among all subtypes of CCSNe. In particular, H-rich, H-poor, and IIn SNe all have a similar median value in logEW(H$\alpha$) $\sim$ 1.6, which is significantly larger than 1.1 measured for Type Ia SNe. The cumulative distribution of dusty SNe spans a significantly narrow range of logEW(H$\alpha$) between Ia and CCSNe, with a median value of 1.25, roughly 0.15 index higher and 0.4 index lower compared to that of Type Ia and CCSNe, respectively. The statistics are also tabulated in Table~\ref{tab:median4types}. 

The differences of the distribution for each SN type combinations are also presented quantitatively in the KS matrix, and the $p$-value of the tests express the statistical significance of two distributions being drawn from the same parent sample. We choose a confidence level of 95\%, that is, we consider that a $p-$value $\leq$ 0.05 as indicating a statistically significant difference between these two distributions. Therefore, the zero KS $p$-value between Ia and all the normal CCSN types suggests a statistically significant difference between them. The other comparisons between all CCSN distributions including the dusty CCSNe, do not show statistically significant differences, compared to that \cite{Pessi2023}\ find a statistically significant difference between H-rich and H-poor SNe based on a smaller sample. 

Additionally, we find that there is an increasing divergence between H-rich SNe and other CCSNe in EW(H$\alpha$) distribution above the median. Plus the higher EW(H$\alpha$) value of H-rich SNe on average, this may indicate more massive stars explode with their hydrogen-envelope retained which is conflict with our understanding on massive star evolution \citep{Smith2011,Eldridge2013}. Can more massive stars explode with their hydrogen envelope retained and give rise to a H-rich type II SNe? We expect more other observations to confirm this. But the measurement of young stellar population age based on EW(H$\alpha$) is debated which will be discussed more in detail in Section \ref{subsec:SFR}. 

\begin{figure}[ht!]
\centering
\includegraphics[width=0.49\textwidth]{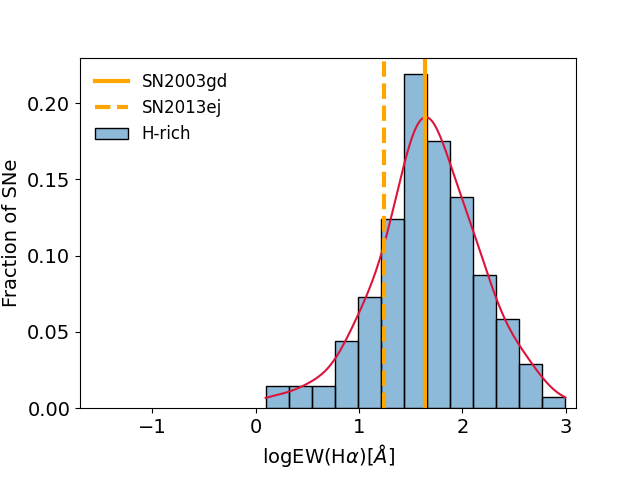}
\includegraphics[width=0.49\textwidth]{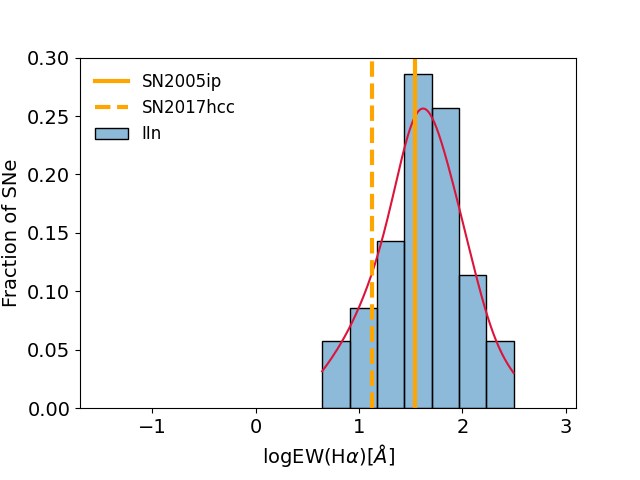}\\
\includegraphics[width=0.49\textwidth]{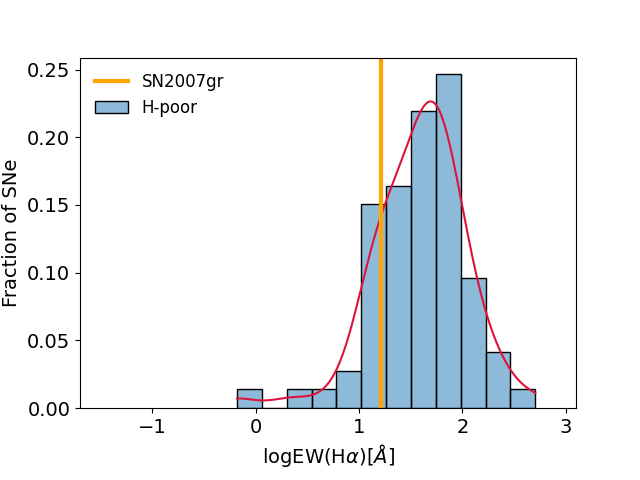}
\includegraphics[width=0.49\textwidth]{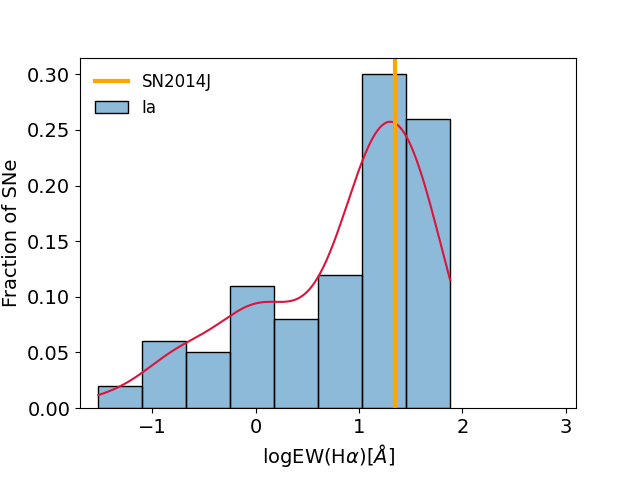}
\caption{
Histograms with kde (the red curves) of H$\alpha$ equivalent width at SN sites in in four types: H-rich (the top left), H-poor (the bottom left), IIn (the top right) and Ia (the bottom right) separately, with cross-matched dusty SN environment EW(H$\alpha$) positions indicated by vertical yellow line(s).
\label{fig:hist_EW_dusty}}
\end{figure}

%\textcolor{blue}{\bf YY: Please check the grammar very carefully if one decide to stick with the current version.}
In Figure \ref{fig:hist_EW_dusty}, we furthermore compare the EW(H$\alpha$) of dusty SN hosts to the normal ones in the four different types, H-rich, H-poor, IIn and Ia respectively. We find all the core-collapse dusty (CC-dusty) SNe host at the lower half region with logEW(H$\alpha$)$<$1.7, compared to their normal counterparts. This may indicate CC-dusty SN progenitor tend to be a less massive population among all the CCSN progenitors. On the other side, the Ia-dusty SN hosts also have their logEW(H$\alpha$)$<$1.7, but they occupy the higher half region of all the normal Ia hosts. It seems that Ia-dusty SNe are more likely discovered at the regions of relatively younger stellar populations. This region is overlapped with that of CCSN hosts. Therefore, CC-dusty and Ia-dusty SN hosts although are on the two sides in their individual groups, they share the same region of logEW(H$\alpha$) values between 1.0 and 1.7 which is a relatively middle region among all SN hosts. We also find CCSNe are rarely located at low EW(H$\alpha$)[$\AA$] $<$0, indicating that their progenitors are more closely tracing star-forming regions. However, the double peak distribution of Ia SNe in EW(H$\alpha$), shows that about 70\% of Ia SNe concentrate on a higher EW(H$\alpha$) region with logEW(H$\alpha$)[$\AA$] $\sim$ 1.5 similar with the distribution of CCSNe, and another 30\% of Ia host at lower EW regions with logEW(H$\alpha$)[$\AA$] $\sim$ 0. This result may be associated with the two different origins of SN Ia \citep{Mannucci2006,Wang2013}. 

\subsection{Star formation rate} \label{subsec:SFR}
\begin{figure}[ht!]
\centering
\includegraphics[width=0.49\columnwidth]{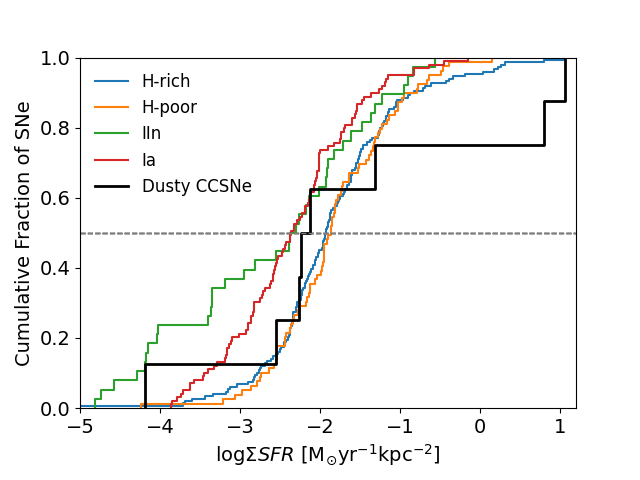}
\includegraphics[width=0.49\columnwidth]{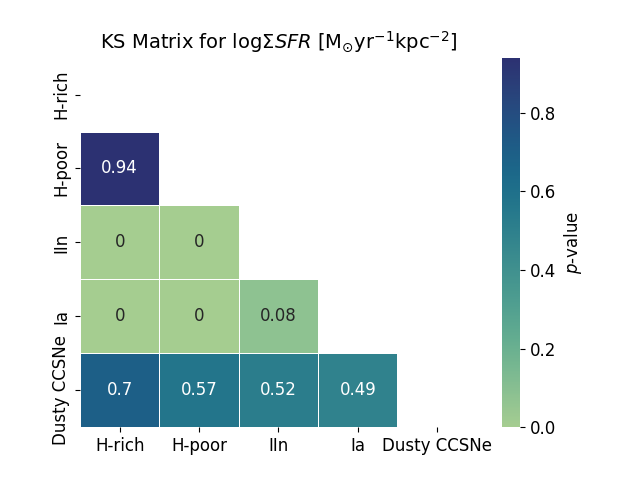}\\
\caption{Left: normalized cumulative distributions of star formation rate intensity at SN sites in four types: H-rich in blue line, H-poor in orange line, IIn in green line and Ia in red line. A dotted horizontal line at 0.5 fraction represents the median value of the distributions. Right: KS statistic matrix for each combination of SN types.
\label{fig:hist_SFR}}
\end{figure}

Figure \ref{fig:hist_SFR} presents the cumulative distributions of SFR intensity in the SNe host HII regions for different SN subtypes and the KS test of each combination. First of all, the cumulative distributions for H-rich and H-poor SNe sites agree remarkably well, which are distinct from the distribution of IIn and Ia. This deviation decreases as the cumulative fraction increase. When the cumulative fraction reaches the median, H-rich and H-poor SN hosts median value of log$\Sigma_{\rm SFR}$[\msolar yr$^{-1}$kpc$^{-2}$] $\sim$ -1.90 which is much higher than those of IIn and Ia with the median value of log$\Sigma_{\rm SFR}$[\msolar yr$^{-1}$kpc$^{-2}$] $\sim$ -2.35. Besides, the comparison between the distributions of Ia and IIn groups displays an overall good agreement above their median values but a gradually increased diversity towards lower cumulative fractions. For the dusty SNe, its distribution lies within the range defined by the span of the normal SN groups below the median, but it is gradually separated out from the normal groups as the cumulative fraction exceeds the median. It seems that dusty SNe have a larger proportion appearing in higher SFR regions, compared to normal types. With the resulting KS test $p$-value between dusty SNe and the normal ones, the difference in SFR distributions are not statistically significant. The $p$-values of zero between IIn and Ia with other CCSNe, suggest a statistically significant difference between these distributions. 

\begin{figure}[ht!]
\centering
\includegraphics[width=0.49\textwidth]{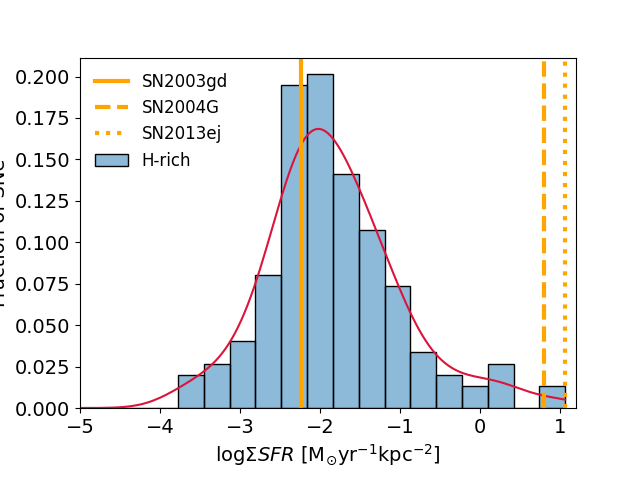}
\includegraphics[width=0.49\textwidth]{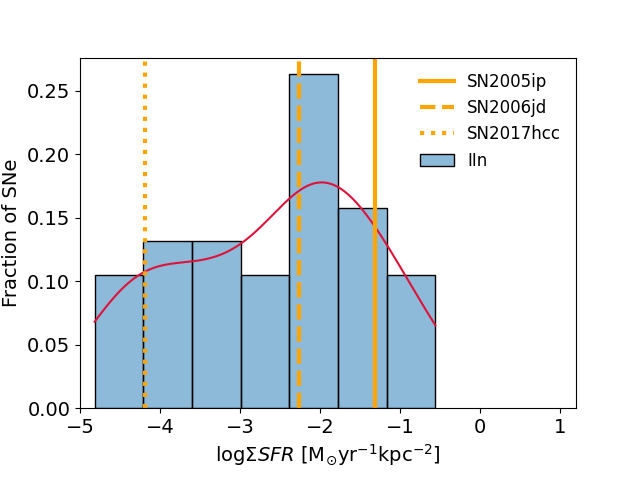}
\includegraphics[width=0.49\textwidth]{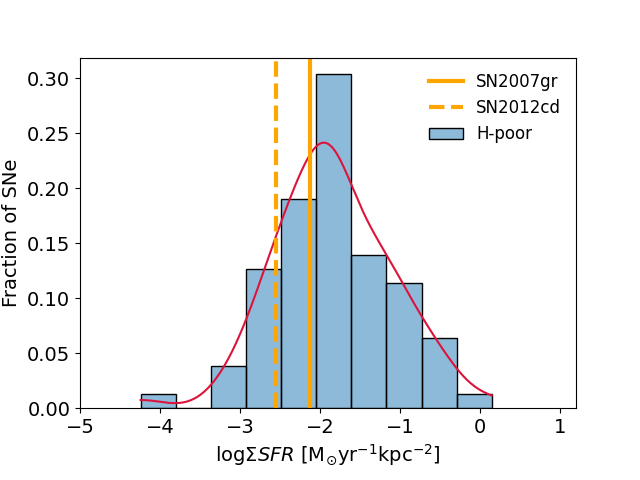}
\includegraphics[width=0.49\textwidth]{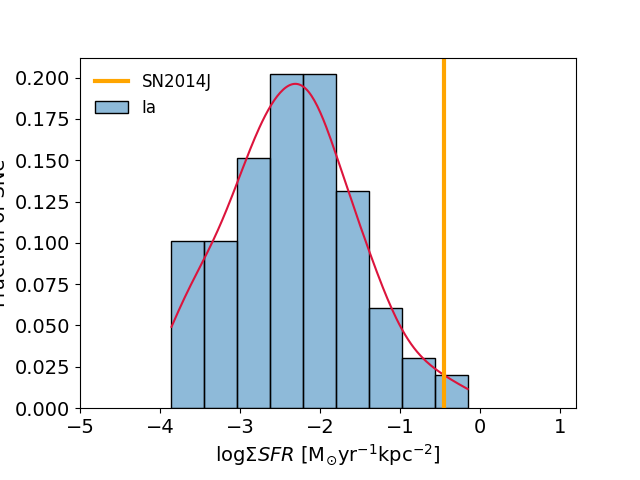}
\caption{Histograms with kde (the red curves) of star formation rate intensity at SN sites in four types: H-rich (the top left), H-poor (the bottom left), IIn (the top right), and Ia (the bottom right) separately, with dusty SN environment star formation rate intensity positions indicated by vertical yellow line(s). 
\label{fig:hist_SFR_dusty}}
\end{figure}

As a whole, H-rich and H-poor SNe are more correlated with ongoing massive star formation than Ia and IIn SNe. In particular, type IIn SNe are more likely to explode at extremely low SFR region. There are more than 20\% of IIn SNe located at the region with log$\Sigma_{\rm SFR}
$[\msolar yr$^{-1}$kpc$^{-2}$] $<$ -4, and this is even much larger than that of Ia SNe. For the dusty SNe, they share the same typical range of SFR with other SN types, but about 20\% of them are likely embedded in regions that probably exhibit the highest star-forming rates. In Figure \ref{fig:hist_SFR_dusty}, the SFR positions of dusty SN hosts in four particular types are presented, and we find it is the H-rich dusty SN hosts that contribute more to the high SFR regions, followed by Ia and IIn. Other dusty SNe that have been measured for relatively higher SFR include Type IIn SN\,2005ip and Type Ia SN\,2014J, as shown in the upper-right and lower-right panels of Figure~\ref{fig:hist_SFR_dusty}, respectively. The lower-left panel suggests that all the H-poor dusty SNe are located at the lower half side of the SFR distribution with log$\Sigma_{\rm SFR}$[\msolar yr$^{-1}$kpc$^{-2}$] $<$ -2. The distribution of IIn SNe in SFR seem to show a similar double-peak feature as the distribution of Ia in EW(H$\alpha$). \cite{Galbany2018} also find two clear components of young and old stellar populations at SNe IIn locations. But this feature does not show in the EW(H$\alpha$) distribution of IIn, and we also do not find a corresponding feature in the SFR distribution of Ia. 

%\textcolor{blue}{\bf YY: do you mean: it would be necessary to investigate the physical processes described by these two parameters.}
To explain these inconsistencies between EW(H$\alpha$) and SFR distributions, it would be necessary to investigate the physical processes described by these two parameters. The star formation rate intensity log$\Sigma_{\rm SFR}$ here is measured from the extinction-corrected H${\rm \alpha}$ flux contributed by young massive stars, tracing the ongoing SFR. EW(H${\rm \alpha}$) is measured by the strength of the line relative to the continuum, which in fact can be an indicator of the strength of the ongoing SFR compared with the past SFR, which reduces with time if no new stars are created. However, considering the size of HII region segregation can be as large as hundreds of parsec in radius, the continuous star formation bursts in the same site are more reasonable, compared to a single instantaneous star formation history. Therefore, the contamination from older stellar populations together with effect of binary-star evolution and photo-leakage can bring uncertainties to the estimation of stellar population age \citep{Eldridge2017,Xiao2018,Sun2021}.   % How to explain these difference? how other studies look like? 

\subsection{Oxygen abundance} \label{subsec:OH}

\begin{figure}[ht!]
\centering
\includegraphics[width=0.49\textwidth]{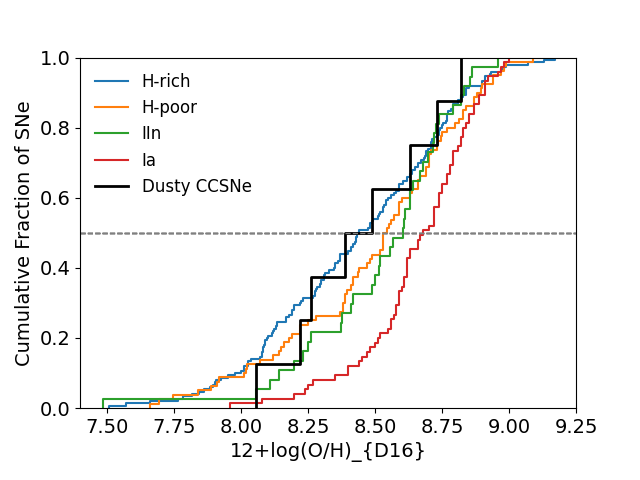}
\includegraphics[width=0.49\textwidth]{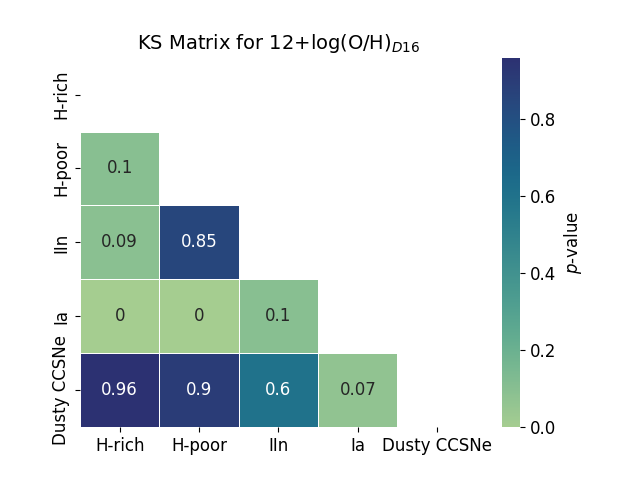}\\
\caption{Left: normalized cumulative distributions of oxygen abundance at SN sites in four types: H-rich in blue line, H-poor in orange line, IIn in green line and Ia in red line. A dotted horizontal line at 0.5 fraction represents the median value of the distributions. Right: KS statistic matrix for each combination of SN types. 
\label{fig:hist_OH}}
\end{figure}

The cumulative distributions of oxygen abundance of the SN host HII regions for different SN subtypes and the KS test for each combination, are present in Figure \ref{fig:hist_OH}. The four normal SN types have distinct median values in their distributions, with the SN Ia distribution having the largest median followed by IIn, H-poor, and H-rich SN hosts. This descending order of oxygen abundance is copied in their median and averaged value as reported in Table \ref{tab:median4types}. When dusty SN hosts are compared to the normal ones, we note that dusty SNe hosts have a relative lower oxygen abundance than the normal types on average, at a similar level of H-rich and H-poor SN hosts. The cumulative distribution of oxygen abundance for dusty SNe are located at the same region as normal ones spanned. Additionally, the KS test values show no statistically significant difference between dusty SNe and the normal ones in oxygen abundance. The $p$-values of zero for Ia and H-rich and H-poor SNe indicate a statistically significant different distribution in Ia SN population which is toward the part of higher oxygen abundance. 

\begin{figure}[ht!]
\includegraphics[width=0.49\textwidth]{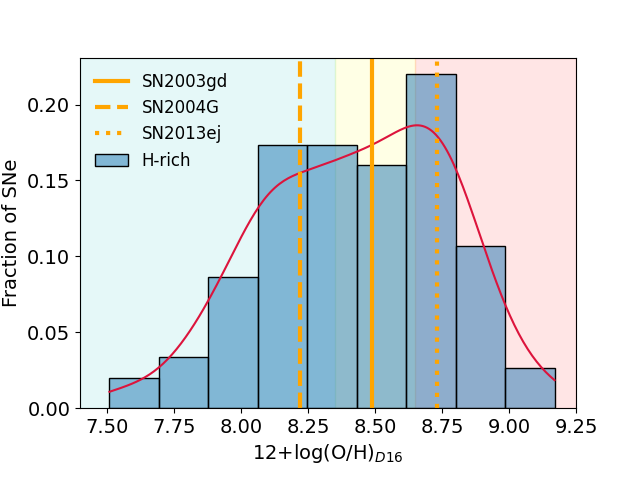}
\includegraphics[width=0.49\textwidth]{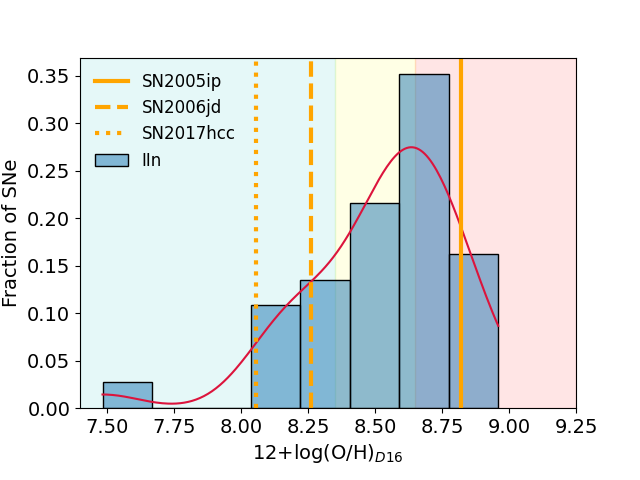}
\includegraphics[width=0.49\textwidth]{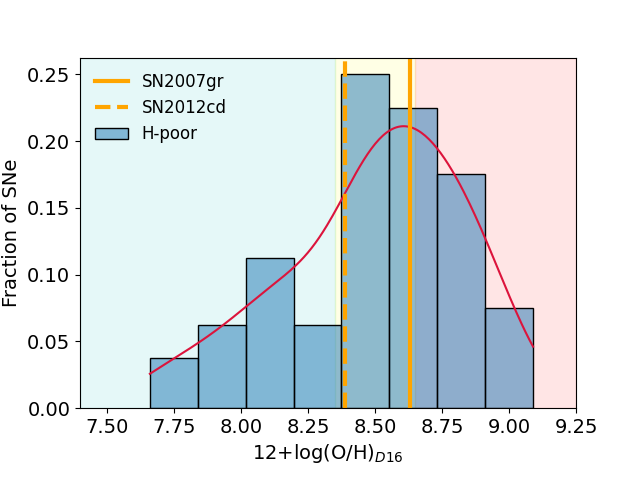}
\includegraphics[width=0.49\textwidth]{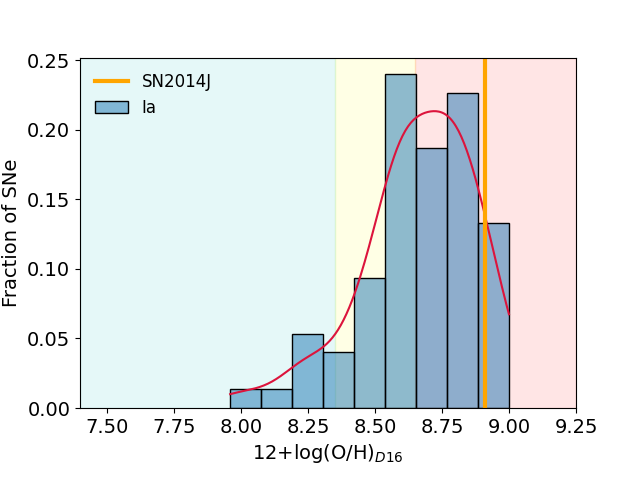}
\caption{Histograms with kde (the red curves) of the oxygen abundance at SN sites in four types: H-rich (the top left), H-poor (the bottom left), IIn (the top right), and Ia (the bottom right) separately, with the dusty SN environment oxygen abundance positions indicated by vertical yellow line(s). We mark the oxygen abundance in three bins (limits at 8.35 and 8.65): subsolar in light blue region, solar in yellow region, and oversolar in pink region.
\label{fig:hist_OH_dusty}}
\end{figure}

In Figure \ref{fig:hist_OH_dusty}, we furthermore compare the oxygen abundance of dusty SN hosts to the normal counterparts in the four types, H-rich, H-poor, IIn, and Ia respectively. We find that the oxygen abundance of dusty SNe scattered a lot from subsolar (12+log(O/H)$_{\rm D16} < $ 8.35) to oversolar (12+log(O/H)$_{\rm D16} >$ 8.65) metallicity. The three H-rich dusty SNe happened to occur in oversolar, solar and subsolar metallicity regions, which may indicate the occurrence of H-rich dusty SNe is less dependent on metallicity. The IIn-dusty SNe prefer the subsolar environment compared to other dusty SN types. We can only find H-poor dusty SNe at solar metallicity environments. The only one Ia-dusty SNe in our sample is located at over-solar metallicity regions. But these results are based on the current sample with fewer number of dusty SNe in each type. We expect more accurate analysis to be carried out based on a more extended sample of dusty SNe in further work.

\subsection{Extinction} \label{subsec:EBV}
\begin{figure}[ht!]
\centering
\includegraphics[width=0.49\textwidth]{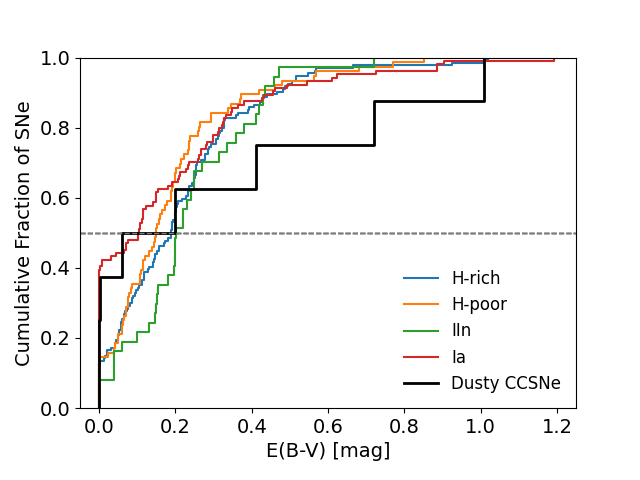}
\includegraphics[width=0.49\textwidth]{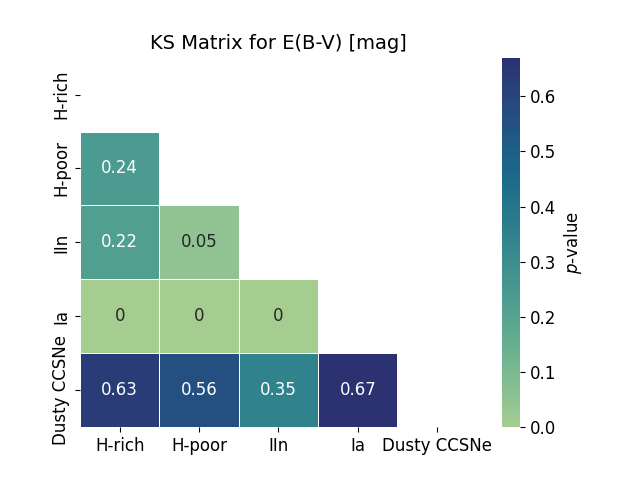}\\
\caption{Left: normalized cumulative distributions of E(B-V) at SN sites in four types: Hydrogen-rich in blue line, Hydrogen-poor in orange line, IIn in green line and Ia in red line. A dotted horizontal line at 0.5 fraction represent the median value of the distributions.Right: KS statistic matrix for each combination of SN types.
\label{fig:hist_EBV}}
\end{figure}

In Figure \ref{fig:hist_EBV} we show the cumulative distributions for the SN host extinction in different SN types and report the KS matrix for each combination of them. First of all, we note that CCSN hosts have larger E(B-V) value than Ia hosts in both median and average, which is also found in \cite{Galbany2017}. It seems that there exits an decreasing order in E(B-V) value from IIn, to H-rich, H-poor and finally to Ia. As for dusty SNe, they happened to occur at the highest extinction regions beyond all the normal ones with E(B-V) up to 0.2 in median and 0.3 in average. The statistical result of these SN host extinction can be found in Table \ref{tab:median4types}. We also find above 40\% of Ia and dusty SN hosts have extremely low and even zero extinction. Compared to the high E(B-V) value of dusty SN hosts on average and median, it seems that the dusty SNe can occur and produce dust in various environments regardless of the level of local inter-stellar medium (ISM). These low extinction regions for other SN types take less than 15\%. Going above the median value into the relatively high extinction regions, the four normal SN types have similar distributions of extinction with E(B-V) varied from 0.2 to 0.6 mag, but dusty SNe are separated out form the main group. This deviation between dusty SNe and the normal types results in about 30\% of dusty SNe occurred at high extinction environments with E(B-V)$>$0.6 mag. This result seems to be consistent with the fact that pre-existing CSM makes normal SNe mid-IR luminous. However, the dust contributing to the extinction of SN hosts are not those produced during the mass-loss history of progenitors, considering that the measured dust radius is much less than the HII region segmentation \citep{Galbany2018}. The extinction of SN hosts is usually thought to be from the formation cloud of the parent stellar population. The ISM existing in the SN host regions can be an extra dust component, which can be heated to observed warm dust in mid-IR by SNe or/and surrounding massive stars when local extinction is high \citep{Foley2014,Yang2017,Bulla2018}. 

\begin{figure}[ht!]
\centering
\includegraphics[width=0.49\textwidth]{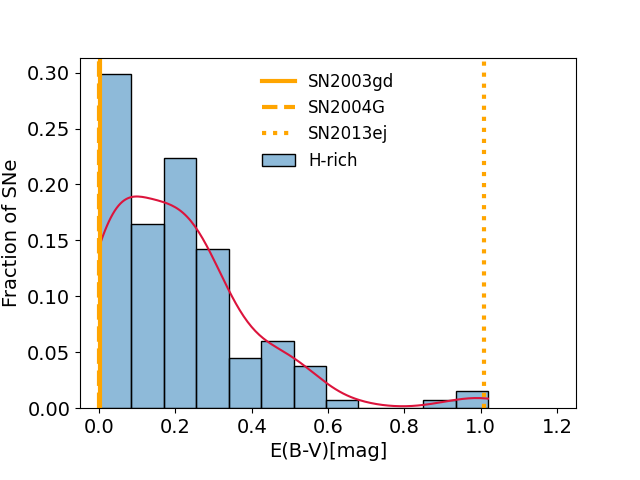}
\includegraphics[width=0.49\textwidth]{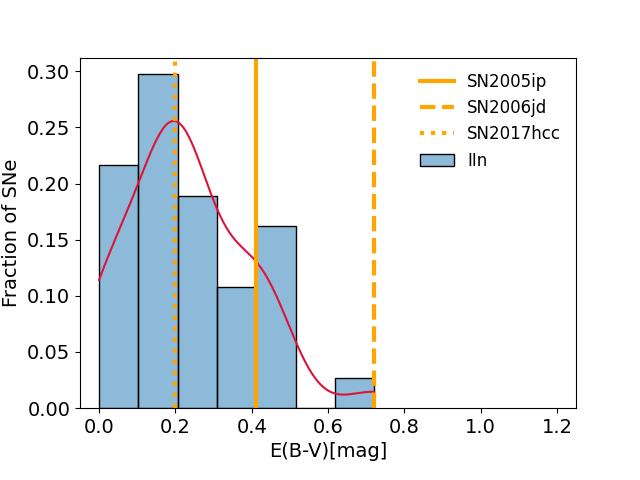}\\
\includegraphics[width=0.49\textwidth]{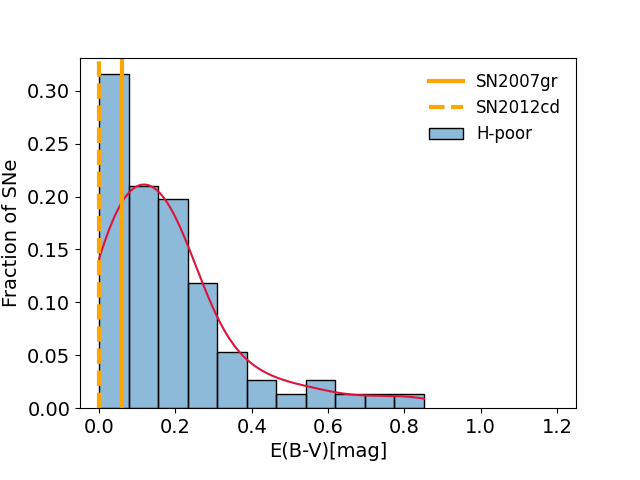}
\includegraphics[width=0.49\textwidth]{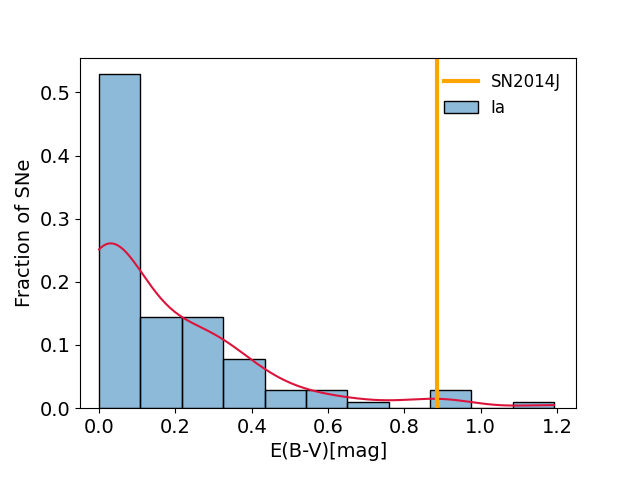}
\caption{Histograms with kde (the red curves)  of E(B-V) at SN sites in four types: H-rich (the top left), H-poor (the bottom left), IIn (the top right) and Ia (the bottom right) separately, with the dusty SN environment E(B-V) positions indicated by vertical yellow line(s). 
\label{fig:hist_EBV_dusty}}
\end{figure}

Then, from the KS matrix we find there are no statistically significant difference between dusty SNe and the normal types in the distributions of E(B-V). The $p$-value of zero appears at the comparison between Ia and all other CCSNe, suggesting a statistically significant difference between these types. We also note the distribution of IIn and H-poor SNe in E(B-V) can be statistically significantly different, with $p$-value of 0.05. In addition, figure \ref{fig:hist_EBV_dusty} presents how different SN types are attribute to the extinction of dusty SN environment in . Given E(B-V) = 0.2 as the lower limit of the high extinction region, we find H-rich, IIn, and Ia are more likely to be discovered in high extinction environments, compared to the fact that all H-poor SNe are located at low extinction environments. However, we cannot ignore the diversity of dusty SN host extinction in all SN types, which can vary from 0 to 1.0 mag. 

\section{Correlating dust properties with environment properties} \label{sec:dependence}
The intensive mass-loss from the SN progenitor systems and dramatic SN explosion itself with the amount of metal-rich ejecta released, reconstruct the environment of star-forming regions and play an important role as dust producer in the Universe \citep{Gall2011,Fox2013}. The origin and heating mechanism of the dust, however, are not always obvious as the dust may be newly formed or pre-existing in the circumstellar medium (CSM) \citep{Pozzo2004,Mattila2008,Smith2009}. Mid-IR SEDs of SNe span the peak of the thermal emission from warm dust marked by a late-time mid-IR excess and can place useful constraints on the dust properties \citep{Szalai2013,Szalai2019}. The statistical analysis of different SN subclasses in \cite{Szalai2019} reveals that subtypes of SNe tend to fill their own regions of phase space in dust parameters (dust luminosity, dust temperature, dust mass) and this seems to correlate with their different progenitor properties \citep{Szalai2019,Szalai2021}. 

%\textcolor{orange}{I am going to Dinner. And figures 9 to 11 will be updated using the new fitting data listed in Table 1 and new result from observations soon. }

The environmental study of SNe can provide good constraints on the progenitor systems confirmed by the literature listed in the introduction part. So in this section, we investigate if there exist any correlations between the SN dust properties with the continuous environmental parameters that can be responsible for it. 

\begin{figure}[ht!]
\centering
\includegraphics[width=0.49\textwidth]{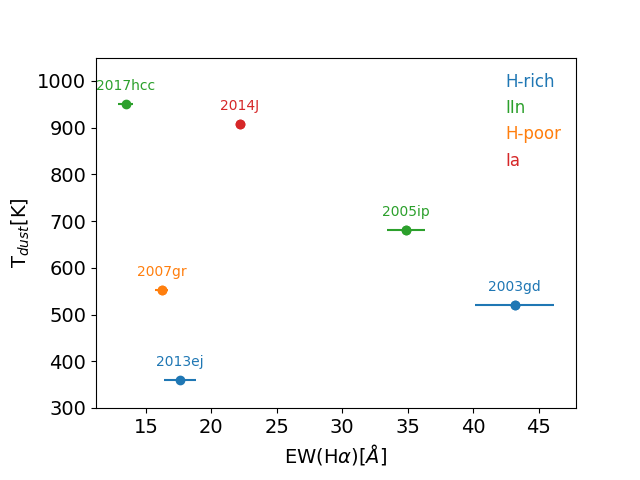}
\includegraphics[width=0.49\textwidth]{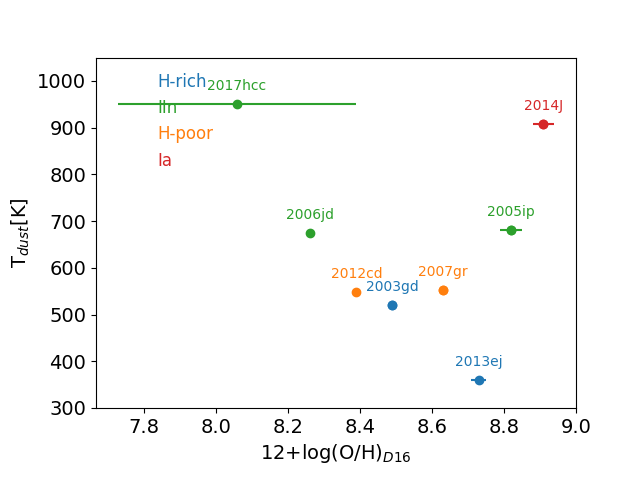}\\
\includegraphics[width=0.49\textwidth]{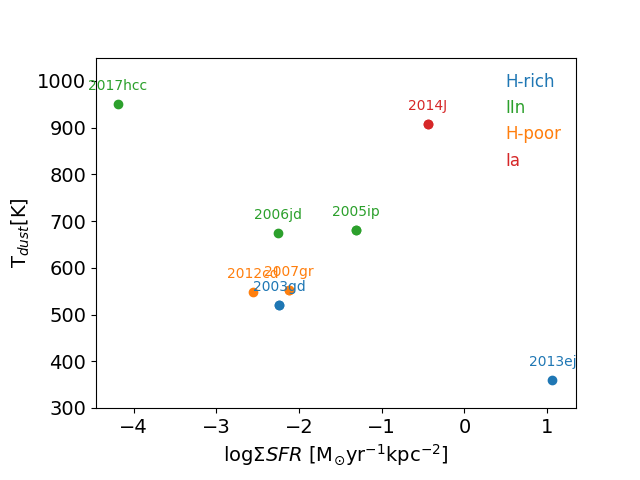}
\includegraphics[width=0.49\textwidth]{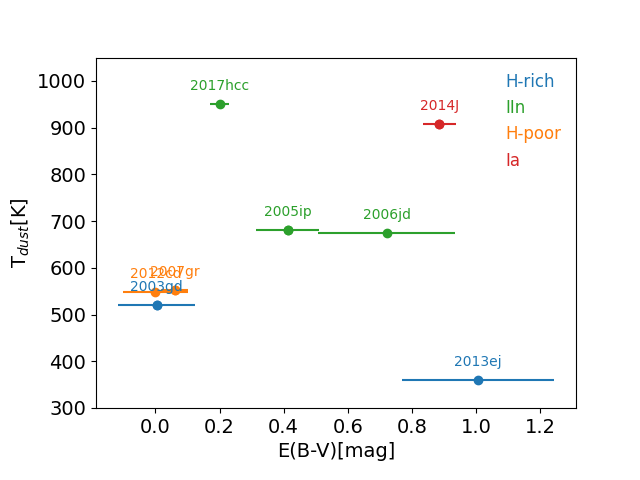}
\caption{The correlation between the dust temperature T$_{\rm dust}$ and the four environmental properties as EW(H$\alpha$) on the top left, oxygen abundance on the top right, Star formation rate intensity on the bottom left and E(B-V) on the bottom right. 
\label{fig:Correlation_Tdust}}
\end{figure}

\begin{figure}[ht!]
\centering
\includegraphics[width=0.49\textwidth]{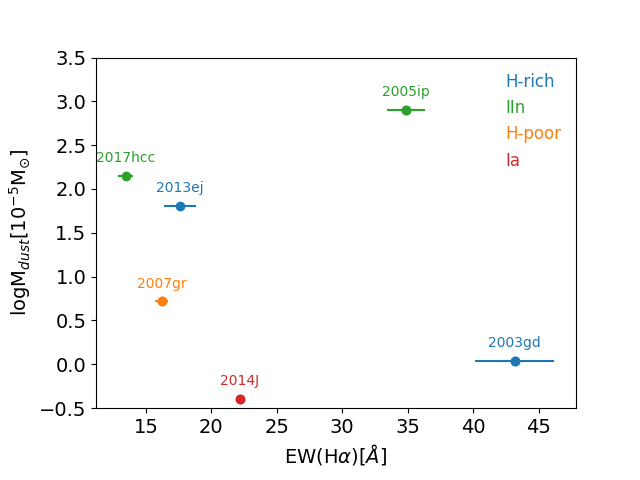}
\includegraphics[width=0.49\textwidth]{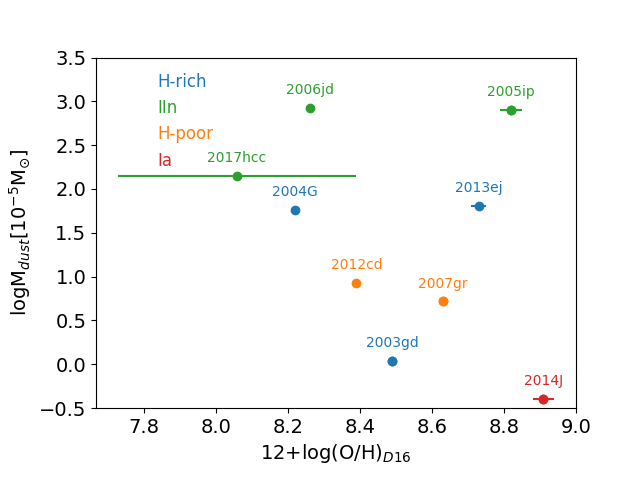}\\
\includegraphics[width=0.49\textwidth]{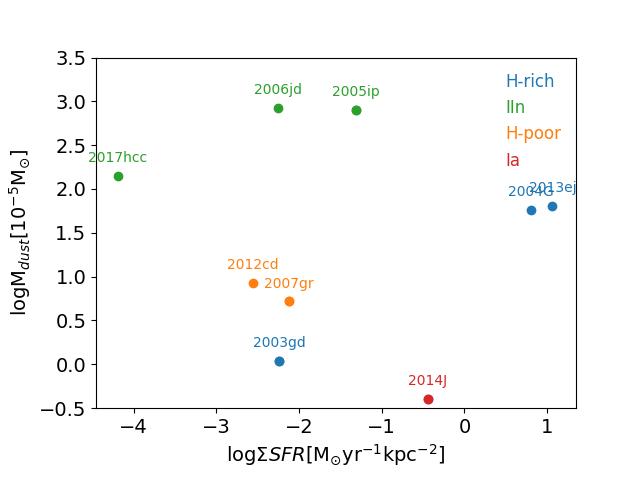}
\includegraphics[width=0.49\textwidth]{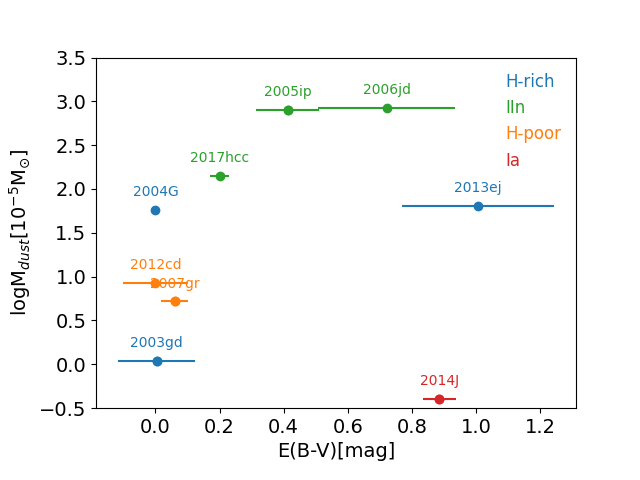}
\caption{The correlation between the dust mass M$_{\rm dust}$ and the four environmental properties as EW(H$\alpha$) on the top left, oxygen abundance on the top right, Star formation rate intensity on the bottom left and E(B-V) on the bottom right. 
\label{fig:Correlation_Mdust}}
\end{figure}

\begin{figure}[ht!]
\centering
\includegraphics[width=0.49\textwidth]{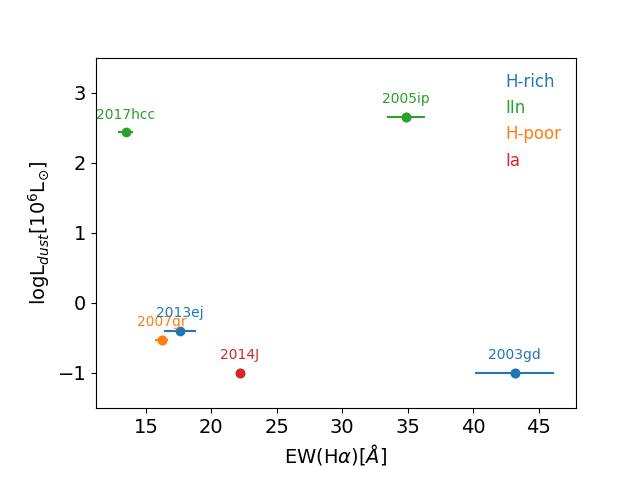}
\includegraphics[width=0.49\textwidth]{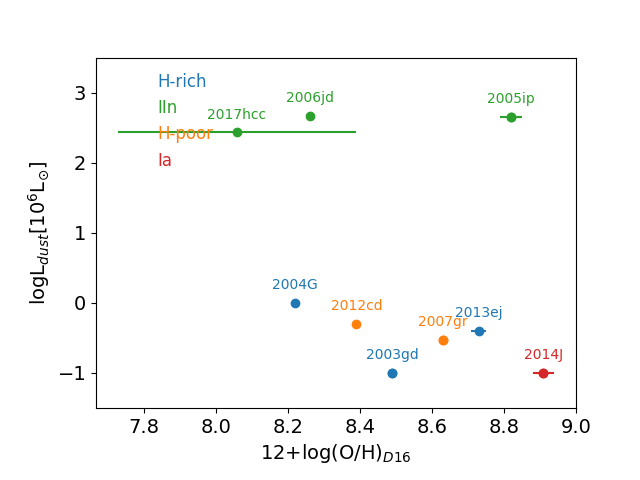}\\
\includegraphics[width=0.49\textwidth]{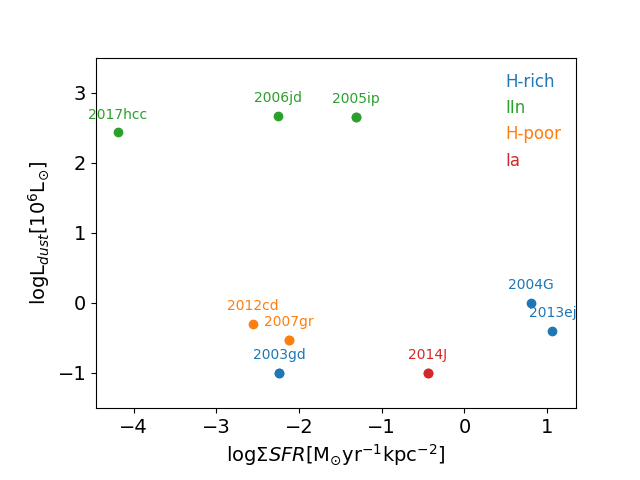}
\includegraphics[width=0.49\textwidth]{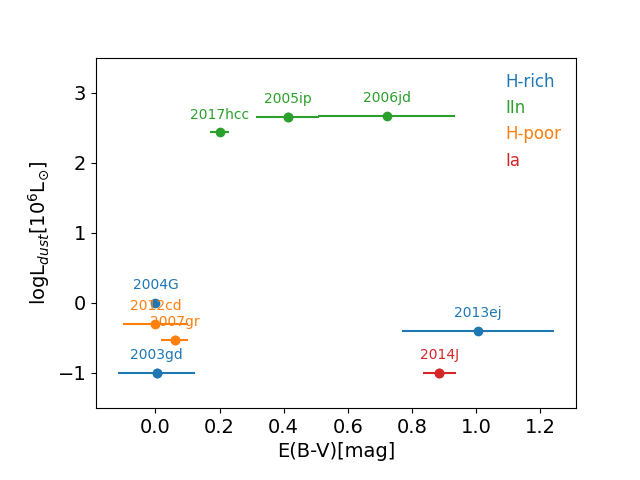}
\caption{The correlation between the dust luminosity L$_{\rm dust}$ and the four environmental properties as EW(H$\alpha$) on the top left, oxygen abundance on the top right, Star formation rate intensity on the bottom left and E(B-V) on the bottom right. 
\label{fig:Correlation_Ldust}}
\end{figure}

The correlation of dust temperature, mass, and luminosity with environmental parameters are presented in Figures~\ref{fig:Correlation_Tdust}, \ref{fig:Correlation_Mdust} and \ref{fig:Correlation_Ldust} respectively. We find the dusty SNe as a whole scattered a lot and any correlation fit failed across the parameter phase. The dusty SNe are more likely to group in the parameter space by their subtypes. Therefore, we look at them separately by their types. Firstly, due to we have only one Ia SN 2014J, we cannot analyze the relationship between its dust parameters and environmental parameters. 

Then, type IIn SNe is the group with the highest similar dust temperature, mass, and luminosity, and they show larger variations in EW(H$\alpha$), oxygen abundance, and $\Sigma$SFR. This is why we cannot work out the correlation between their dust and environmental properties. However, the SN host extinction E(B-V) can group the IIn-dusty SNe best and this may indicate the effect of the similar pre-SNe dusty environment for the measured dust properties. Comparatively, the H-poor dusty SNe are grouped with similar environmental parameters and similar dusty parameters. This is because we have only two of them and their dusty parameters are all based on the fitting model not directly from observations. But their similar low value of E(B-V) may indicate that it is the similar progenitor systems of H-poor dusty SNe that build their dusty CSM environments and measured dust, regardless of their low host extinction.

Finally, the H-rich SNe show the most diversities in both dust and environmental parameters. The metallicity and SFR seem to affect the production of dust mass. The environments with lower metallicity and high SFR can benefit for dust production, leading to higher dust luminosity and mass. However, we note that the current result is not a confident correlation due to the small sample size of dusty SNe, with only three H-rich dusty SNe included. The findings need further confirmation as the sample of the local environment of the dusty SNe sample expands.

\section{Conclusions and future prospects} \label{sec:conclusions}
In this work, we collected 395 SN sites with IFU observations, which is the largest and most diverse in terms of SN types in literature, so that we can update the result of the environmental studies of typical SNe. Furthermore, we compared the environmental properties of the dusty SNe with their normal family SN populations in terms of four subgroups: H-rich, H-poor, IIn, and Ia, which covers all major SN subtypes. We find that the dusty SNe in general do not have a preference for particular environments and scatter a lot in different subtypes. We also investigated the correlation of the dust parameters with those environmental parameters to study how the host environment can affect the detected warm dust from mid-IR excess of SNe at later times. Our main conclusions can be summarized as follows. 

\begin{enumerate}
  \item we find there is an increasing divergence between H-rich SNe and other CCSNe in EW(H$\alpha$) distribution above the median. Plus the higher EW(H$\alpha$) value of H-rich SNe on average, this may indicate more massive stars explode with their hydrogen-envelope retained which is conflict with our understanding on massive star evolution \citep{Smith2011,Eldridge2013}. The relative rate of massive stars explode with their hydrogen envelope retained to those lost is debated. We expect more other observations to confirm it. 
  
  \item Type IIn SNe may even prefer the extremely low SFR region than Ia SNe. We find there are more than 20\% of IIn SNe located at the region of the log$\Sigma_{\rm SFR} <$-4 where we rarely see Ia SNe and other types. This finding throws a doubt on our understanding of the progenitor systems and explosion mechanisms of IIn so far. 
  
  \item The dusty SNe do not closely trace the star forming region. Compared to normal types, dusty SNe have a larger proportion appearing in higher SFR regions.
  
  %However, if some exploded in the high SFR regions, they are more likely to be H-rich dusty SNe.
  
  \item The oxygen abundance of dusty SNe scattered a lot from subsolar to oversolar metallicity, which may indicate the occurrence of dusty SNe is less dependent on metallicity. Although it seems that dusty SNe in different types have some preference in metallicity, this result is based on the current sample with fewer number of dusty SNe in each type and is required to be confirmed by a more extended sample of dusty SNe in further work.
  
  \item The occurrence of dusty SNe is also less affected by their host extinction. We find about 40\% of dusty SN hosts have extremely low and even zero E(B-V), but there are also about 30\% of them located at high extinction environments with E(B-V)$>$0.6 mag. However, it is required to be considered that the ISM existing in the SN host regions, can be an extra dust component which can be heated to observed warm dust in mid-IR by SNe or/and surrounding massive stars when local extinction is high \citep{Foley2014,Yang2017,Bulla2018}. 

  \item We find the dusty SNe as a whole scattered a lot and show little correlation between dusty and environmental parameters. The dusty SNe are more like to group in the parameter space based on their subtypes.

  \item For H-rich dusty SNe, the environments with lower metallicity and high SFR can benefit for the production of warm dust, leading to higher dust luminosity, and mass. 

\end{enumerate}

However, We note that limited by the small sample size of dusty SNe presented in this study, one should remain cautious about the inferred correlations between various parameters, which require further confirmation with more data. This is the goal of a specific ongoing work combining the IFS data with a larger mid-IR sample using {\it WISE} detection (Xiao et al. in prep). We expect enlarge the sample size for all SN subtypes, which is more essential to study the correction between dust parameters and environmental parameters and bring us more confirmed result for the effect of the SN host environments on the warm dust detected from the mid-IR excess at later times. 

Additionally, we remark that limited by the IFS data which is designed to study galaxies not only for SN host galaxies. There are considerable amount of nearby SNe without corresponding IFS data. Some interesting studies using images that are not IFS data, analyze the environmental properties of the SN hosts and provide constraints on progenitor populations in an alternative way \citep{Ransome2022,Li2023}. The studies open an alternative path accessing to SN environmental study, and encourage the diversity in observations. 

Finally, we expect more observational data and theoretical models to improve our understanding of dusty SNe and their environments. The photometry and spectrometry data from {\it JWST} can may offer unprecedented opportunity to resolve the details of the immediate environment of the dusty SNe, trace the evolution of dust from SN to SN remnant phase, and test more precise dust models with different components \cite{Shahbandeh2023,Zsiros2023,Sarangi2022}.

%% Also note that the akcnowlodgment environment does not support long amounts of text. If you have a lot of people and institutions to acknowledge, do not use this command. Instead, create a new \section{Acknowledgments}.

\begin{acknowledgments}
We thank all the people that have made this paper. LX is thankful for the support from National Natural Science Foundation of China (grant No. 12103050 ), Advanced Talents Incubation Program of the Hebei University, and Midwest Universities Comprehensive Strength Promotion project. Y.Y. appreciates the generous financial support provided to the supernova group at U.C. Berkeley by Gary and Cynthia Bengier, Clark and Sharon Winslow, Sanford Robertson, and numerous other donors. T.P. acknowledges the support by ANID through the Beca Doctorado Nacional 202221222222.
\end{acknowledgments}


\vspace{5mm}
\facilities{{\it Spitzer}, {\it WISE}, {\it CAO:3.5m(PMAS/PPak)}, {\it VLT/MUSE}, {\it APO:2.5m(SDSS)}}


\software{astropy \citep{2013A&A...558A..33A,2018AJ....156..123A},  
          numpy \citep{2013RMxAA..49..137F}, 
          Matplotlib \citep{1996A&AS..117..393B},
          Seaborns \citep{Waskom2021},
          Photutils \citep{larry_bradley_2023_7946442},
          SFFT \citep{sfft_zenodo}}



\appendix

\section{The mid-IR photometry of SN 2014J}\label{sec:2014J-WISE}
\begin{figure*}[ht!]
\begin{minipage}{0.45\columnwidth}
\includegraphics[width=\columnwidth]{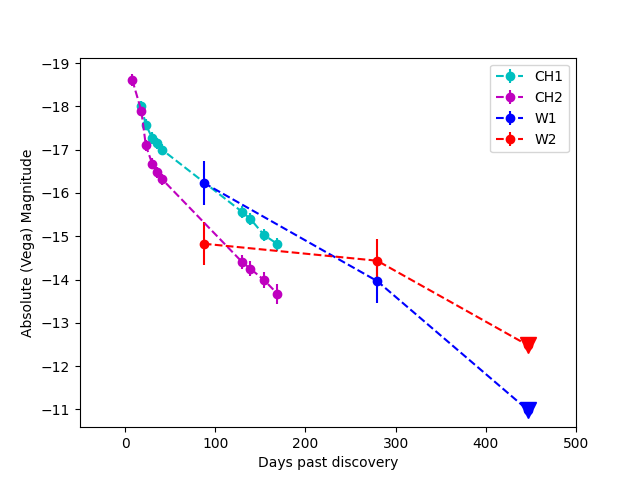}
\caption{The mid-IR light curve of SN 2014J from {\it WISE} (W1=3.4$\mu$m, the blue points, and W2=4.6$\mu$m, the blue points), compared to the detection from {\it Spitzer} (CH1=3.6$\mu$m,CH2=4.5$\mu$m). The triangles mark the upper limits of the detection at that epoch. 
\label{fig:mirLC}}
\end{minipage}
\end{figure*}

We present the mid-IR photometry data of SN 2014J from {\it Spitzer} and {\it WISE} in this section, and the later-time mid-IR detection is used to estimate the dust properties of SN 2014J in Section \ref{sec:sample}. The {\it Spitzer} photometry data of SN 2014J is collected from \cite{Szalai2019}. We made the mid-IR photometry of SN 2014J using the time-resolved {\it WISE/NEOWISE} coadded images \citep{Meisner2018}. For all the post-explosion images, we performed image subtraction with SFFT \citep{SFFT}. Then we obtained the aperture photometry of each epoch with Photutils \citep{larry_bradley_2023_7946442}. Figure \ref{fig:mirLC} presents the mid-IR light curve of SN 2014J reduced from {\it WISE} image, compared to early-time light curve from {\it Spitzer}. 

%\textcolor{red}{LH: more details on data processing for Spitzer and WISE?}. 

\begin{deluxetable*}{ccc}
%\begin{minipage}{0.45\columnwidth}
\tablecaption{The mid-IR photometry of SN 2014J from {\it WISE} (W1=3.4$\mu$m, W2=4.6$\mu$m).  \label{tab:LC-data-2014J}}
% \tablenum{1}
\tablecolumns{3}
\tablehead{
\colhead{ Epoch } & \multicolumn{2}{c}{Absolute mag.}  \\
\cline{2-3} 
\colhead{days} & \colhead{W1 (3.4$\mu$m)} & \colhead{W2 (4.6$\mu$m)} 
}
\startdata
\tableline
87 & -16.23$\pm$0.12 & -14.83$\pm$0.13  \\
280 & -13.97$\pm$0.17 & -14.44$\pm$0.13  \\
447 &  $<$ -10.98 & $<$ -12.48  \\
\tableline
\enddata
%\end{minipage}
\end{deluxetable*} 

\section{The linear regression of dust evolution}\label{sec:linear_dust}
\begin{figure}[ht!]
\includegraphics[width=0.35\columnwidth]{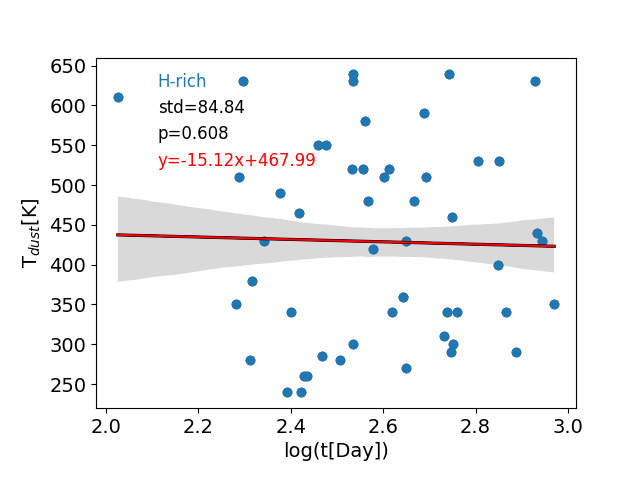}
\includegraphics[width=0.35\columnwidth]{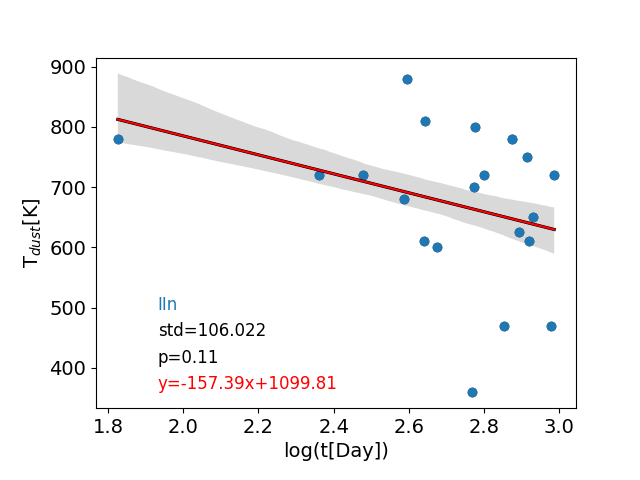}\\
\includegraphics[width=0.35\columnwidth]{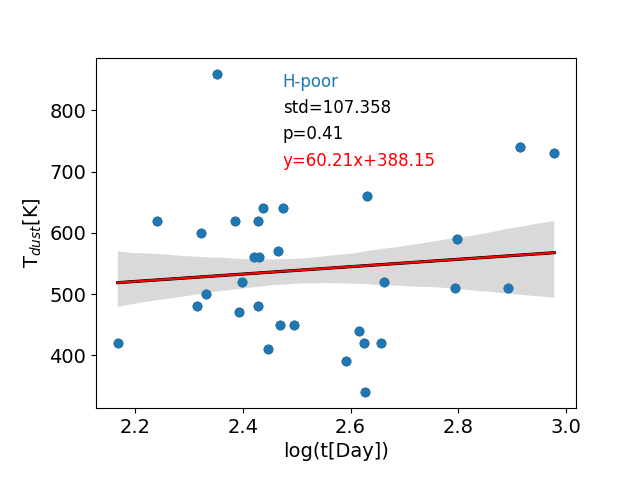}
\includegraphics[width=0.35\columnwidth]{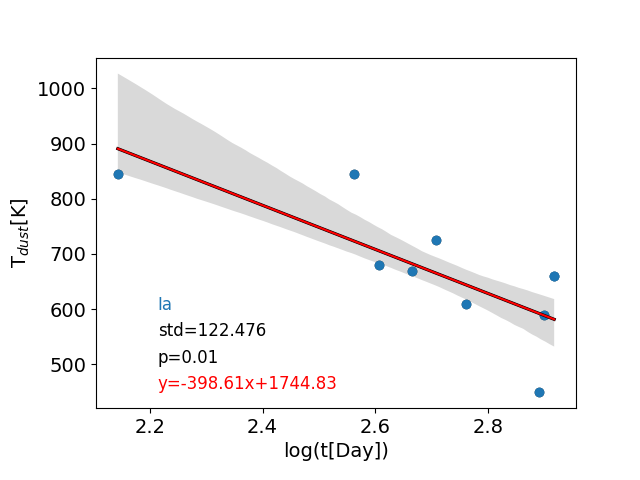}
\caption{The linear regression of dust temperature evolution for SNe (the blue dots) inferred from \cite{Szalai2019}. The red line is fitting result and the gray area marks the 1-$\sigma$ uncertainties. 
\label{fig:inter_Tdust}}
\end{figure}

In Figure \ref{fig:inter_Tdust} we present how dust properties evolve with time in SNe inferred from \cite{Szalai2019}, and a linear regression is given with 1-$\sigma$ uncertainties. The data using for the linear regression is carefully selected. Firstly, we made a cutoff in time at 1000 days, because observed dust are most likely to appear between 200 and 600 days and data completeness decreases dramatically beyond 1000days. Secondly, we removed the SNe that their dust parameters are measurement limits, and the data with certain measurements are used to make the linear regression. Because the luminous type IIn and the type 2008S-like are very distinct with each other in mid-IR properties \citep{Szalai2019}, it is impossible using a mixed data of these two types to make a linear regression. Considered the SNe IIn in our sample listed in Table \ref{tab:sndata} are all luminous SNe IIn, we made the linear regression using the data with the 2008S-like SNe removed. 

We consider the fitting results for any particular subtype of SNe plausible when a criterion of regression $<0.5$ is reached. we use the linear regression to estimate the dust properties for the SNe without later-time observations in Table \ref{tab:sndata}. We note the dust temperature in H-rich and H-poor increase with time, and this results in the dust mass decreases with time. This trend is dominant by the mid-IR observations during timescale $<800$ day past SNe. After 600 days the 1-$\sigma$ confidence interval increase due to insufficient data supply. 

\begin{figure}[ht!]
\includegraphics[width=0.35\columnwidth]{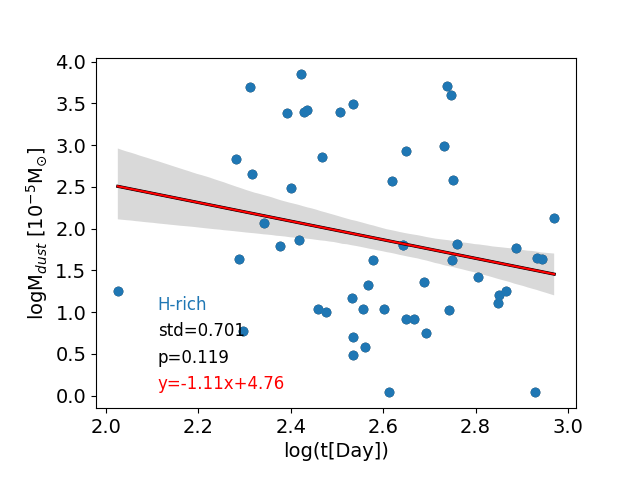}
\includegraphics[width=0.35\columnwidth]{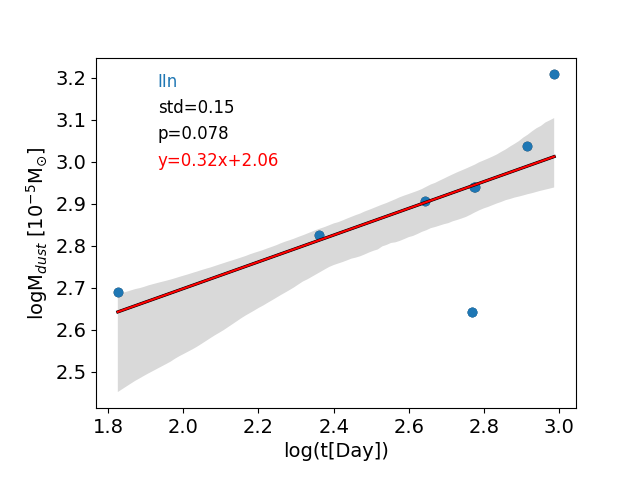}\\
\includegraphics[width=0.35\columnwidth]{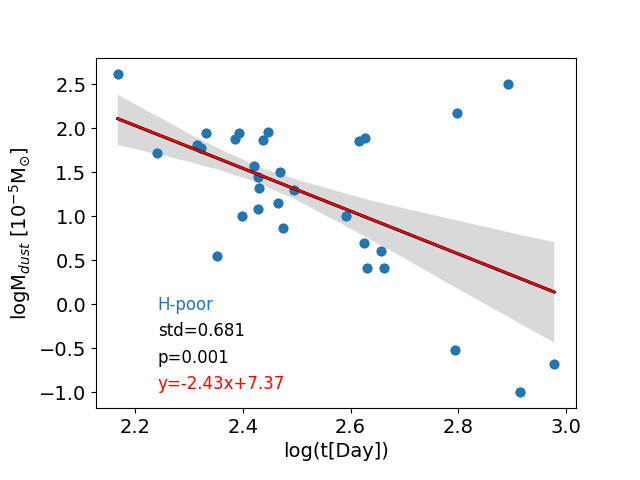}
\includegraphics[width=0.35\columnwidth]{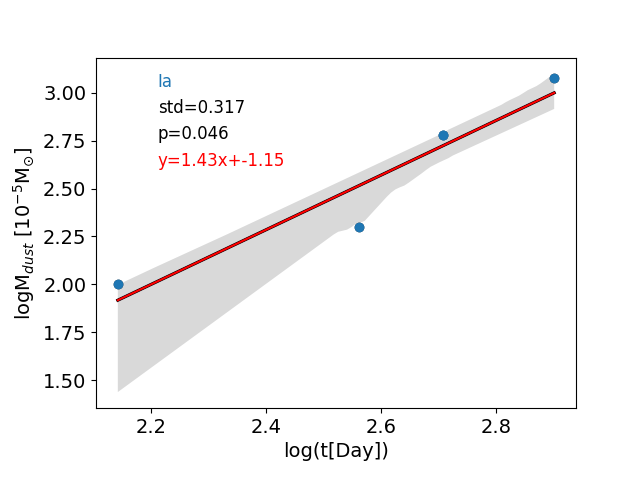}
\caption{The linear regression of dust mass evolution for SNe (the blue dots) inferred from \cite{Szalai2019}. The red line is fitting result and the gray area marks the 1-$\sigma$ uncertainties.
\label{fig:inter_Mdust}}
\end{figure}

\begin{figure}[ht!]
\includegraphics[width=0.35\columnwidth]{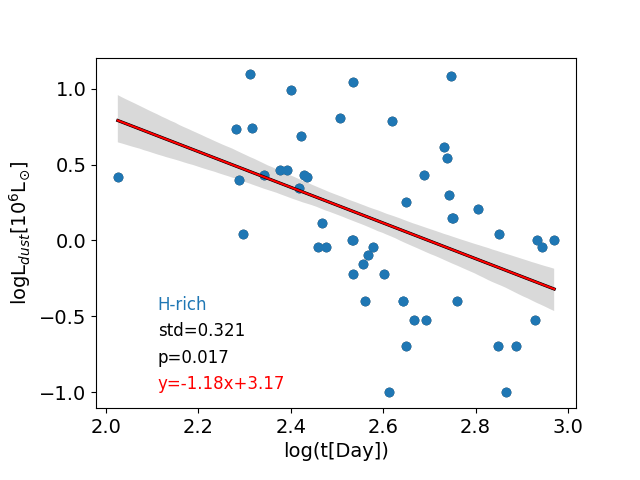}
\includegraphics[width=0.35\columnwidth]{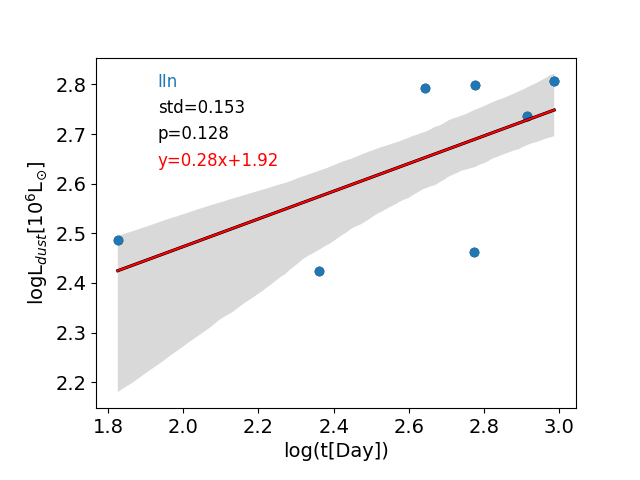}\\
\includegraphics[width=0.35\columnwidth]{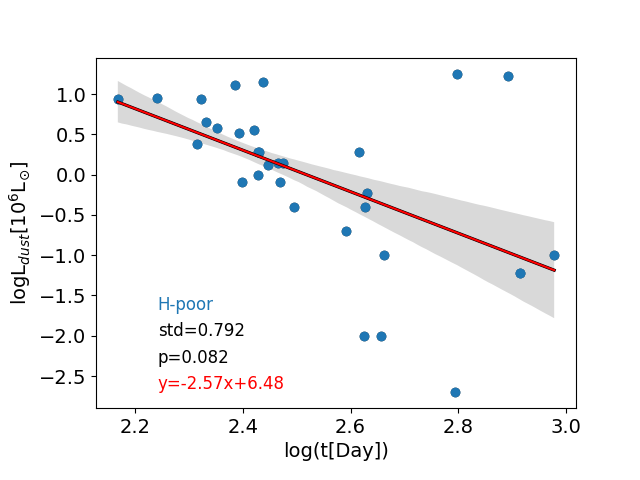}
\includegraphics[width=0.35\columnwidth]{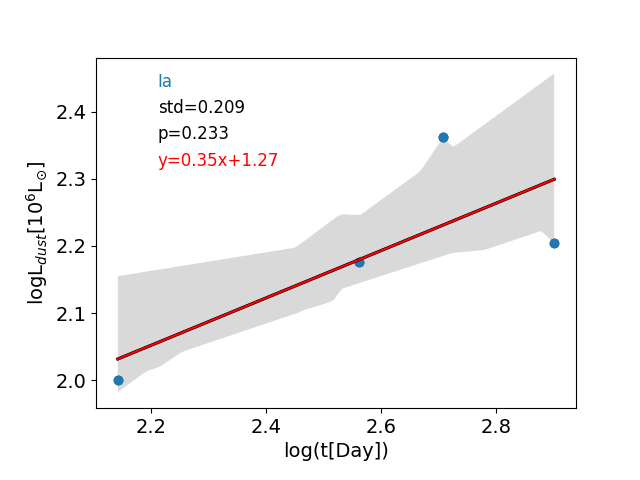}
\caption{The linear regression of dust luminosity evolution for SNe (the blue dots) inferred from \cite{Szalai2019}. The red line is fitting result and the gray area marks the 1-$\sigma$ uncertainties.
\label{fig:inter_Ldust}}
\end{figure}


\bibliography{sample631}{}
\bibliographystyle{aasjournal}



\end{document}